\documentclass[sigconf]{acmart}

\usepackage[english]{babel}
\usepackage{blindtext}

\usepackage{comment}
\usepackage{algorithmic}
\usepackage{graphicx}
\usepackage{textcomp}
\usepackage{booktabs}
\usepackage{multirow}
\usepackage{xcolor}
\usepackage{wrapfig}
\usepackage{pdfpages}
\usepackage{tabularx}
\usepackage{balance}
\usepackage{float}

\AtBeginDocument{%
  \providecommand\BibTeX{{%
    \normalfont B\kern-0.5em{\scshape i\kern-0.25em b}\kern-0.8em\TeX}}}

\setcopyright{acmcopyright}
\copyrightyear{xxx}
\acmYear{xxx}
\acmDOI{xxx}
\acmConference[ACM MobiCOVID'22]{ACM MobiCOVID '22: xxx}{Seoul}{South Korea}
\acmBooktitle{xxx}
\acmPrice{xxx}
\acmISBN{xxx}

\renewcommand\footnotetextcopyrightpermission[1]{} 
\setcopyright{none}
\acmDOI{}
\acmISBN{}
\acmPrice{}

\begin{document}

\title[Comparison of Popular Video Conferencing Apps]{Comparison of Popular Video Conferencing Apps Using Client-side Measurements on Different Backhaul Networks
}



\author{Rohan Kumar \hspace{0.5mm} Dhruv Nagpal\\ Vinayak Naik}
\email{[f20181013, f20180095, naik]@goa.bits-pilani.ac.in}
\affiliation{%
  \institution{BITS Pilani, Goa}
  \country{India}
}

\author{Dipanjan Chakraborty}
\email{dipanjan@hyderabad.bits-pilani.ac.in}
\affiliation{%
  \institution{BITS Pilani, Hyderabad}
  \country{India}
}
\renewcommand{\shortauthors}{Rohan Kumar et al.}

\begin{abstract}
Video conferencing platforms have been appropriated during the COVID-19 pandemic for different purposes, including classroom teaching. However, the platforms are not designed for many of these objectives. 
When users, like educationists, select a platform, it is unclear which platform will perform better given the same network and hardware resources to meet the required Quality of Experience (QoE). Similarly, when developers design a new video conferencing platform, they do not have clear guidelines for making design choices given the QoE requirements.

In this paper, we provide a set of networks and systems measurements, and quantitative user studies to measure the performance of video conferencing apps in terms of both, Quality of Service (QoS) and QoE. Using those metrics, we measure the performance of Google Meet, Microsoft Teams, and Zoom, which are three popular platforms in education and business.
We find a substantial difference in how the three apps treat video and audio streams.
We see that their choice of treatment affects their consumption of hardware resources.
Our quantitative user studies confirm the findings of our quantitative measurements.
While each platform has its benefits, we find that no app is ideal.
A user can choose a suitable platform depending on which of the following, audio, video, or network bandwidth, 
CPU, or memory are more important.

\end{abstract}

\maketitle


\section{Introduction}

When the COVID-19 pandemic hit, countries worldwide went into strict lockdown, and schools, universities, offices, and places of business closed down. Video conferencing platforms like Google Meet, Microsoft Teams, and Zoom were appropriated in different domains like classroom education, healthcare, family functions, corporate events, meetings, and shopping, for people to continue functioning. However, the video conferencing platforms were not envisioned to be used in scenarios where the usage and network infrastructure are very diverse in terms of devices employed and bandwidth. With the continuing cycle of COVID waves, many of the platforms will likely be continued to be used for different purposes, including classroom education. However, vital domains such as school education have been badly hit during the COVID-19 pandemic, especially in developing countries, because of several socio-cultural factors, including the affordability of devices and network bandwidth \cite{cscw-blind}. In this work, we focus on the technological factors affecting the quality of school and university education during the COVID-19 pandemic. We conduct Quality of Service (QoS) experiments through client-side network measurements on three popular video conferencing platforms, namely, Google Meet, Microsoft Teams, and Zoom, in an ecologically valid scenario for classroom education, under different network conditions with varying modes of operation within the apps. We also conduct Quality of Experience (QoE) experiments through quantitative user studies over the same platforms subjected to the same network and operational variations. Our work serves to inform educationists in developing countries on choosing a platform to continue conducting classroom education in online modes. In addition, our work tells designers of platforms in prioritizing different aspects for the education domain.


In the absence of access to server-side measurements, we conduct client-side measurements for the QoE experiments to determine several network characteristics, like upload and download payload sizes and the Inter-Packet Arrival Times (IPAT). We also make quantitative comparisons between the audio and video qualities at the sender and the receiver sides. 
We conduct quantitative user studies to see how the network 
and hardware
usage of the various apps affect the user experience. These insights, we believe, will empower policymakers and educationists to choose a platform for their needs. On the other hand, it will inform developers in low-resource contexts about which characteristics or features are essential.

The salient contributions of this paper are as follows.
\begin{enumerate}
    \item We study the network usage at the client-side of three popular video-conferencing platforms and correlate that with video and audio quality to understand whether and how the two are related.
    \item We conduct this study using Google Meet, which is widely used in the education domain, and Microsoft Teams and Zoom, which are commonly used in the corporate space and education.
    \item We quantitatively measure network usage 
    and video-audio quality. Since video-audio quality is also a subjective metric, we quantitatively measure the perceived quality through a user experience study.
    \item We use bandwidth, download payloads, upload payloads, and IPAT (inter-packet arrival times) to measure network characteristics.
    \item We quantitatively measure video characteristics in terms of PSNR (Peak Signal to Noise Ratio) and SSIM (Structural Similarity Index Measure). 
    \item We measure energy in different audio frequencies, the bitrate, and the number of channels for studying the audio characteristics.
    \item We measure these metrics for the three apps on wired broadband over WiFi and 4G mobile Internet connections. These two varieties of connections are the most widely used for online classes during the COVID-19 pandemic.
    \item In our measurements, the bandwidth of wired broadband on optical fiber is roughly about 150 Mbps while that of 4G mobile Internet is about 11 Mbps. We want to see how the platforms behave in terms of their network usage and video-audio quality when presented with different backhaul networks.
    \item We vary the settings of the platforms in terms of microphone and camera as they result in creating different payloads for the network.
\end{enumerate}

\noindent \textbf{Organization of the Paper} The paper is organized as follows. Section \ref{sec:measuring} describes the metrics for quantitative measurements of network usage, 
video quality, and audio quality. All of these count towards the Quality of Service (QoS) class of metrics. We explain the survey that we conduct to measure the quality of video and audio qualitatively, which count towards the class of QoE. In Section \ref{sec:performance}, we analyze the performance of the two apps using the metrics for wired broadband and 4G mobile Internet connections. We compare our work with the existing bodywork in Section \ref{sec:related}. Finally, we conclude in Section \ref{sec:conclusion} and mention the future work in Section \ref{sec:future}.



\section{Measuring the Performance}
In this section, we describe each of the metrics we use and discuss the setup used to collect values of those metrics.

\label{sec:measuring}
\subsection{Measuring Quantitative Performance}
We measure the apps' performance in terms of their network usage at the client-ends, both transmitter, and receiver of the video. 
An advantage of measuring at the client-end is that no special access is required at the server. 
Any user can measure performance without needing any special access to the apps.
The Upload Payload is the total payload in the packets sent from the video source to the server.
The Download Payload is the total payload in the packets sent from the server to the video receiver.
The IPAT is the time difference between any two successive packets at the receiver.
To compare the network performance of both the apps, we measure Upload Payload at the transmitter-end of the video, Download Payload at the receiver-end of the video, and Standard Deviation in IPAT. 
We analyze CPU utilization, memory usage, and battery consumption for the two different networks.

We perform the measurements over a session for each app, lasting for fifteen minutes. Over the session, we play a recorded video of a lecture from a university, which mimics the scenario of streaming live or recorded classes and meetings for which these apps are heavily being used.
In total, we perform twelve different combinations for the session, depending on whether the microphone and camera are switched ON or OFF.
We tabulate these combinations in Table \ref{tab:test-variations}. 
These twelve test combinations give us all the possible configurations of the state of the apps and the accessories.
While the video contains the speaker and the slides, the camera transmits the video at the receiver's end.
Since we need at least one speaker for the video conference, the speaker's video is transmitted in all twelve combinations.
We use Wireshark
to capture sent and received packets. We use Numpy and Pandas Python libraries with the Wireshark packet capture to compute the network metrics.
We use a python script to measure resource consumption of the conferencing apps' processes.
The script uses the psutil~\cite{psutils} library to capture resource consumption characteristics.

We record the sessions using the app's recording feature to measure the video quality and audio quality against the local copy of the video and audio.
There are multiple techniques available to evaluate video quality~\cite{seshadrinathan-2010B}. 
We use PSNR and SSIM to compare the video quality. 
PSNR is a quantitative video quality metric that gives us the inverse of the error between the original and the recorded frames. A higher PSNR indicates better quality.
SSIM is a more complex quantitative metric that considers perceptual quality~\cite{zoran-2010}. 
Its value lies between zero and one, the latter value implying that the two frames are the same.
We use the YUV color encoding to calculate the SSIM and PSNR values. 
`Y' component depicts the brightness, `U' the blue projection, and `V' the red projection~\cite{msu}.
We use Spek~\cite{spek} to compare the audio quality, which gives us energy distribution for different audible frequencies. 
The higher the energy distribution among the frequencies, the better is the audio quality \cite{campbell-2009}. 
We repeat each measurement three times on different days and report an average of those. 

We conduct the measurements over two network configurations - (a) wired broadband networks with the end-hosts connected via a WiFi network and (b) 4G mobile Internet. 
We give details of the configurations of networks and end-hosts in Table \ref{tab:s_n}.
To the extent possible, we keep the configuration the same at sender and receiver.

\begin{table}
\caption{Measurements Performed}\label{tab:test-variations}
  \centering
  \begin{tabular}{c| c c c}
    \toprule
    {\small\textit{Seq No}}
    & {\small \textit{App}}
      & {\small \textit{Mic}}
    & {\small \textit{Camera}} \\
    \midrule
    1 & Google Meet & OFF & OFF \\
    2 & Google Meet & ON & OFF \\
    3 & Google Meet & OFF & ON \\
    4 & Google Meet & ON & ON \\
    5 & MS Teams & OFF & OFF \\
    6 & MS Teams & ON & OFF \\
    7 & MS Teams & OFF & ON \\
    8 & MS Teams & ON & ON \\
    9 & Zoom & OFF & OFF \\
    10 & Zoom & ON & OFF \\
    11 & Zoom & OFF & ON \\
    12 & Zoom & ON & ON \\
    \bottomrule
  \end{tabular}
\end{table}

\begin{table}
\caption{Configuration of network and end-hosts}~\label{tab:s_n}
\centering
\begin{tabular}{ p{0.1\textwidth} p{0.15\textwidth} p{0.15\textwidth}}\\
\toprule
{\small\textit{Participant's Role}}         & {\small\textit{Sender}}          & {\small\textit{Receiver}} \\     
\midrule
CPU                      & Intel i5-8265U           & Intel i5-8250U           \\
\hline
RAM         &16 GB       &16 GB \\
\hline
OS         & Windows 10               & Windows 10               \\
\hline
Broadband Internet Connection & WLAN 802.11ac over 150 Mbps Optical Fiber & WLAN 802.11ac over 150 Mbps Optical Fiber\\
\hline
4G Mobile Internet & 11 Mbps & 10.5 Mbps\\
\hline
Battery & 41 Wh & 41 Wh\\
\hline
Browser & Google Chrome & Google Chrome\\
\bottomrule
\end{tabular}
\end{table}

\subsection{Measuring Qualitative Performance}
We measure qualitative performance to assess the users' video and audio experience. 
We survey to determine the factors that influence qualitative user experience and correlate with the quantitative metrics. 
Once we can establish a correlation, the app developers will improve overall user experience and product performance by focusing on these measurable metrics.

We take help of fifteen participants to evaluate the qualitative performance. 
We ask these survey subjects to view the original video before showing them the same video transmitted over Google Meet, Microsoft Teams, and Zoom. We ask them to gauge differences in the quality of  streamed content in terms of 
Video Quality, Audio Quality, Resolution, Video-Audio Synchronisation, Buffering/Frame Drops, and Lag on a 5-point Likert scale, with one being the worst and five being the best. We randomize the order of the contents across all the subjects.

\section{Performance Analysis} \label{sec:performance}
We analyze the collected quantitative and qualitative metrics and correlate the two.

\subsection{Quantitative Performance over Wired Broadband via WiFi}
\begin{figure}
    \centering
    \includegraphics[scale=0.5]{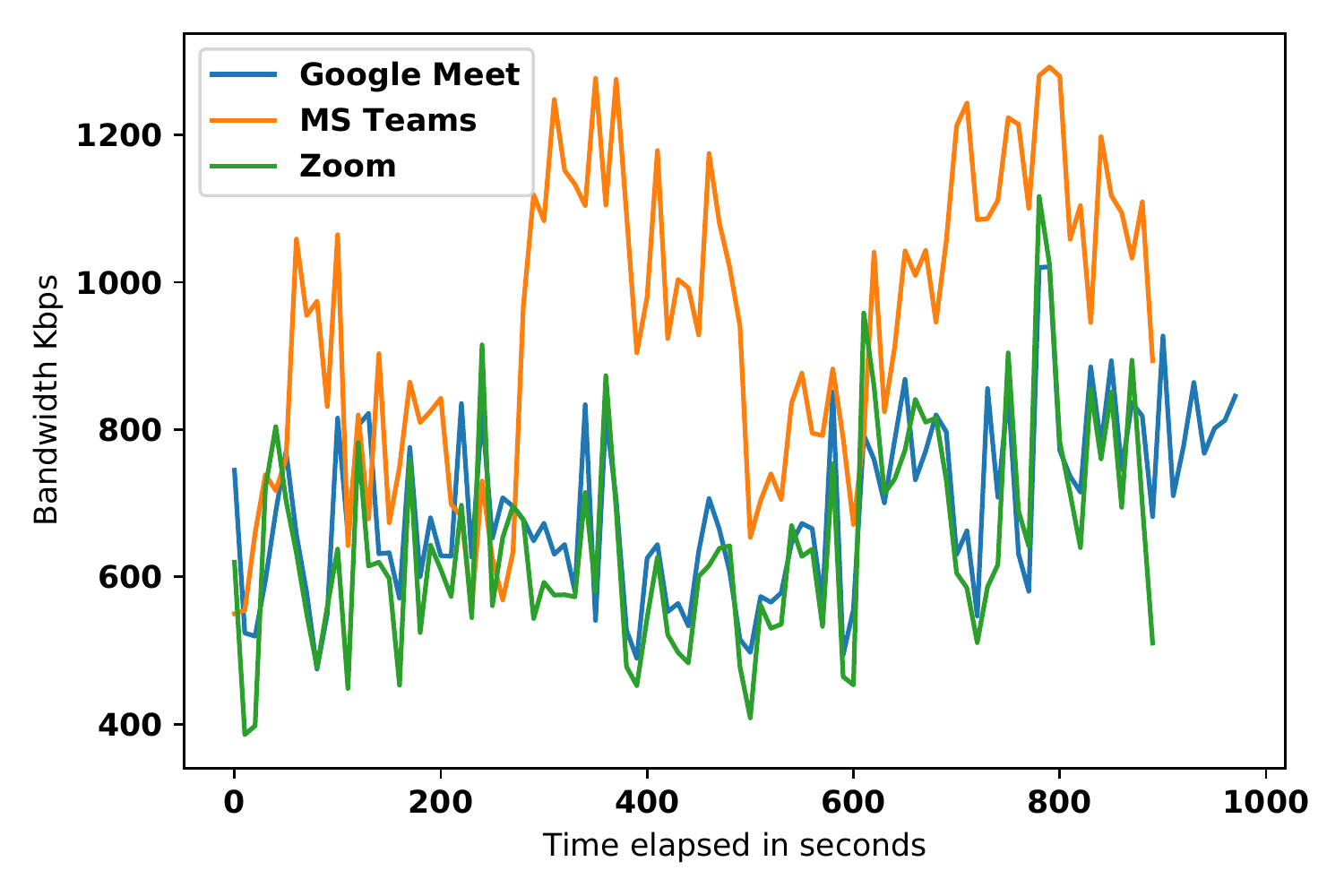}
    \caption{We plot the Download Bandwidth every 10 sec for wired broadband via WiFi. Microsoft Teams uses higher bandwidth. Google Meet and Zoom not only use lower bandwidth, they are also more stable as the two have lower standard deviations. The standard deviation for Google Meet bandwidth is 120.16 Kbps, Microsoft Teams is 231.6 Kbps, and Zoom is 131.21 Kbps.}\label{fig:dl_bw}
\end{figure}

\begin{figure}
    \centering
    \includegraphics[scale=0.5]{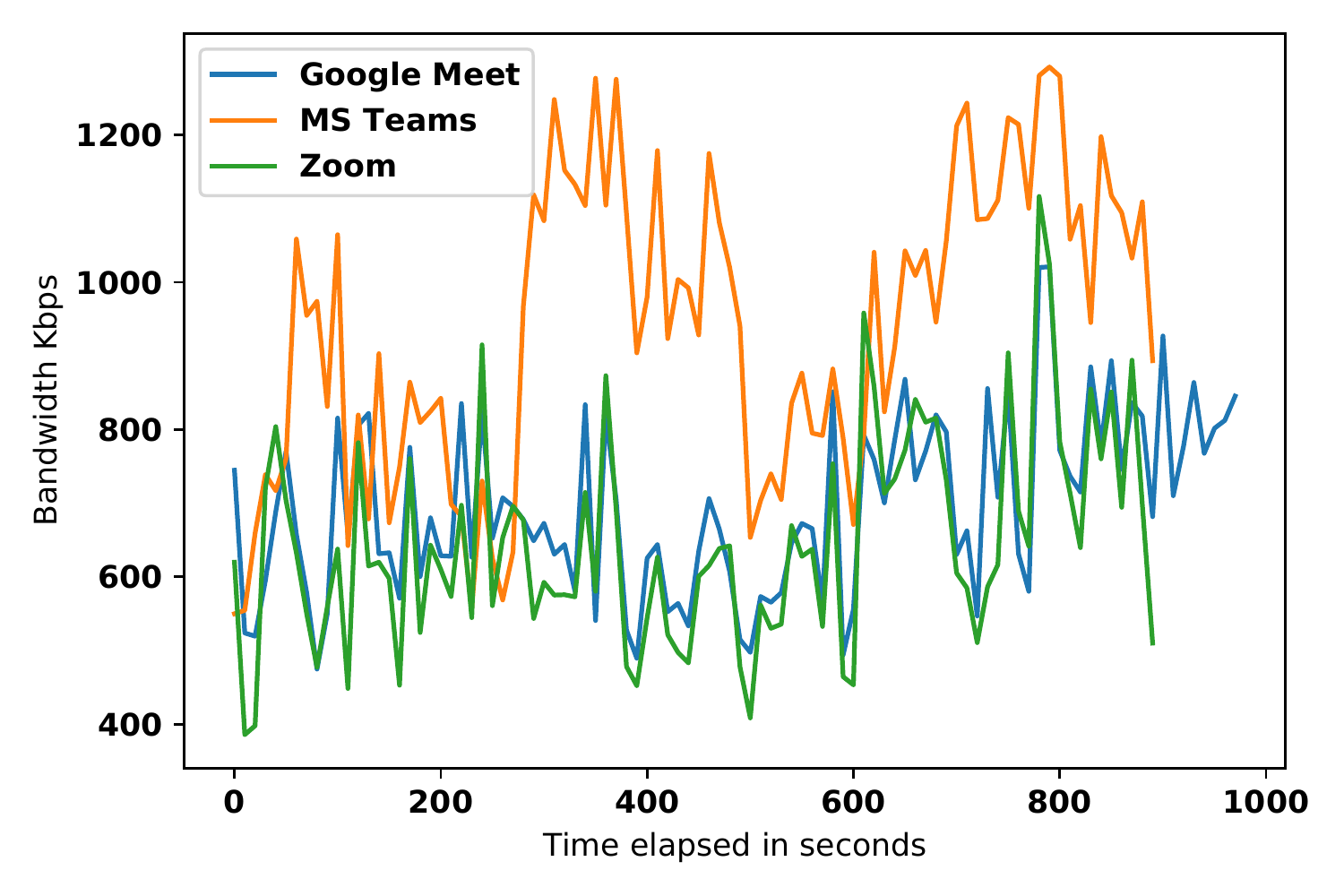}
    \caption{We plot the Upload Bandwidth every 10 sec for the wired broadband via WiFi. Microsoft Teams uses higher bandwidth. Google Meet and Zoom not only use lower bandwidth , they are also more stable as the two have lower standard deviations. The standard deviation for Google Meet bandwidth is 112.44 Kbps, Microsoft Teams is 211.52 Kbps, and Zoom is 117.81 Kbps.}\label{fig:ul_bw}
\end{figure}

In Table~\ref{tab:results_wbb}, we see that Microsoft Teams uses higher Bandwidth and approximately 10\% higher Payloads for all the twelve measurements, which implies that it is sending more data from the sender to the server and from the server to the receiver.
We present a detailed view of Upload Bandwidth and Download Bandwidth when both the mic and the camera are switched ON, in Figures \ref{fig:dl_bw} and \ref{fig:ul_bw}, respectively.
We observe similar plots for all the other seven measurements.
Due to space constraints, we show plots only for one measurement.
The standard deviations in IPAT for Zoom and Google Meet are minor, which is essential for having low jitter~\cite{rao-2019}. 
The standard deviation in IPAT for Microsoft Teams is almost twice that of Google Meet and Zoom when both the mic and the camera are OFF, as seen in Table \ref{tab:results_wbb}. This implies that the packets pertaining to the video being played by the sender are sent irregularly in the case of Microsoft Teams. It will result in a poor perceived quality of the video.
We see in Table \ref{tab:results_wbb} that Zoom has a considerably higher PSNR for all tests, which suggests that the video streaming of Zoom contains minimal noise as compared to Microsoft Teams and Google Meet.
The video quality of Google Meet was reduced by a noticeable amount when the camera was switched ON. A dip indicates this in the SSIM value. 
A low SSIM value when the camera is switched ON suggests that Google Meet compresses the screen-sharing video to a greater extent to compensate for the added payload when the camera is switched ON. The PSNR data of Microsoft Teams and Zoom are higher than Google Meet in all the measurements. On further inspecting Y, U, and V components of SSIM 
in Table \ref{tab:yuv_comparison}
, we observe that the `Y' value for Google Meet is significantly lower than that of Microsoft Teams and Zoom when both the microphone and camera are switched ON, but the `U' and `V' values are comparable. 
A lower `Y' values indicate that Google Meet compromises on the luminescence of the video to save bandwidth when the camera is switched ON.

\begin{table*}
\centering
\caption{Summary of data collected over Wired Broadband over WiFi. Microsoft Teams uses higher data bandwidth than Google Meet and Zoom. However, Zoom delivers better video quality assessment scores. Google Meet and Zoom have a lower standard deviation of IPAT, which provides a smoother experience while viewing.}\label{tab:results_wbb}
\begin{tabular}{@{}lcccccc@{}}
\toprule
\textit{Measurement Type} & \multicolumn{1}{l}{\textit{App}} & \multicolumn{1}{l}{\textit{\begin{tabular}[c]{@{}l@{}}Download \\ Payload (MB)\end{tabular}}} & \multicolumn{1}{l}{\textit{\begin{tabular}[c]{@{}l@{}}Upload \\ Payload (MB)\end{tabular}}} & \multicolumn{1}{l}{\textit{\begin{tabular}[c]{@{}l@{}}$\sigma({IPAT})$\\ (ms)\end{tabular}}} & \multicolumn{1}{l}{\textit{\begin{tabular}[c]{@{}l@{}}PSNR\\ (YUV)\end{tabular}}} & \multicolumn{1}{l}{\textit{\begin{tabular}[c]{@{}l@{}}SSIM\\ (YUV)\end{tabular}}} \\ \midrule
\multirow{2}{*}{Mic OFF Cam OFF} & Google Meet & 54 & 58 & 16.64 & 33.98 & 0.97 \\
 & MS Teams & 69 & 87 & 33.48 & 37.58 & 0.98 \\
 & Zoom & 52 & 55 & 13.74 & 49.45 & 0.99 \\
\hline
\multirow{2}{*}{Mic ON Cam OFF} & Google Meet & 60 & 63 & 12.63 & 33.75 & 0.96 \\
 & MS Teams & 76 & 88 & 12.13 & 37.21 & 0.98 \\
  & Zoom & 61 & 63 & 9.43 & 49.15 & 0.99\\
\hline
\multirow{2}{*}{Mic OFF Cam ON} & Google Meet & 74 & 77 & 8.11 & 32.11 & 0.88 \\
& MS Teams & 83 & 95 & 8.50 & 37.04 & 0.98 \\
& Zoom & 74 & 76 & 7.21 & 48.24 & 0.99\\
\hline
\multirow{2}{*}{Mic ON Cam ON} & Google Meet & 79 & 82 & 8.07 & 31.26 & 0.85 \\
 & MS Teams & 93 & 100 & 8.36 & 36.92 & 0.98 \\
 & Zoom & 77 & 79 & 7.19 & 47.53 & 0.99\\
\bottomrule
\end{tabular}
\end{table*}

\begin{table}
\centering
\caption{The values of Y, U, and V towards the computation of SSIM. We find a significant drop in 'Y' value for Google Meet when the camera is switched ON.}
\label{tab:yuv_comparison}
\begin{tabular}{ccccc}
\toprule
{\small\textit{Test}} {\small\textit{Type}} & {\small\textit{Platform}} & {\small\textit{Y}} & {\small\textit{U}} & {\small\textit{V}} \\
\midrule
\multirow{2}{*}{Mic OFF Cam OFF} & Google Meet & 0.97 & 0.97 & 0.98 \\
 & MS Teams & 0.97 & 0.98 & 0.99 \\
 & Zoom & 0.99 & 0.99 & 0.99\\
\hline
\multirow{2}{*}{Mic ON Cam OFF} & Google Meet & 0.95 & 0.96 & 0.97 \\
 & MS Teams & 0.98 & 0.98 & 0.99 \\
 & Zoom & 0.99 & 0.99 & 0.99\\
\hline
 \multirow{2}{*}{Mic OFF Cam ON} & Google Meet & 0.66 & 0.97 & 0.97 \\
 & MS Teams & 0.97 & 0.99 & 0.99 \\
 & Zoom & 0.99 & 0.99 & 0.99\\
\hline
\multirow{2}{*}{Mic ON Cam ON} & Google Meet & 0.61 & 0.96 & 0.98 \\
 & MS Teams & 0.97 & 0.98 & 0.99 \\
 & Zoom & 0.99 & 0.99 & 0.99\\
\bottomrule
\end{tabular}
\end{table}

\begin{figure}
    \centering
    \includegraphics[scale=0.4]{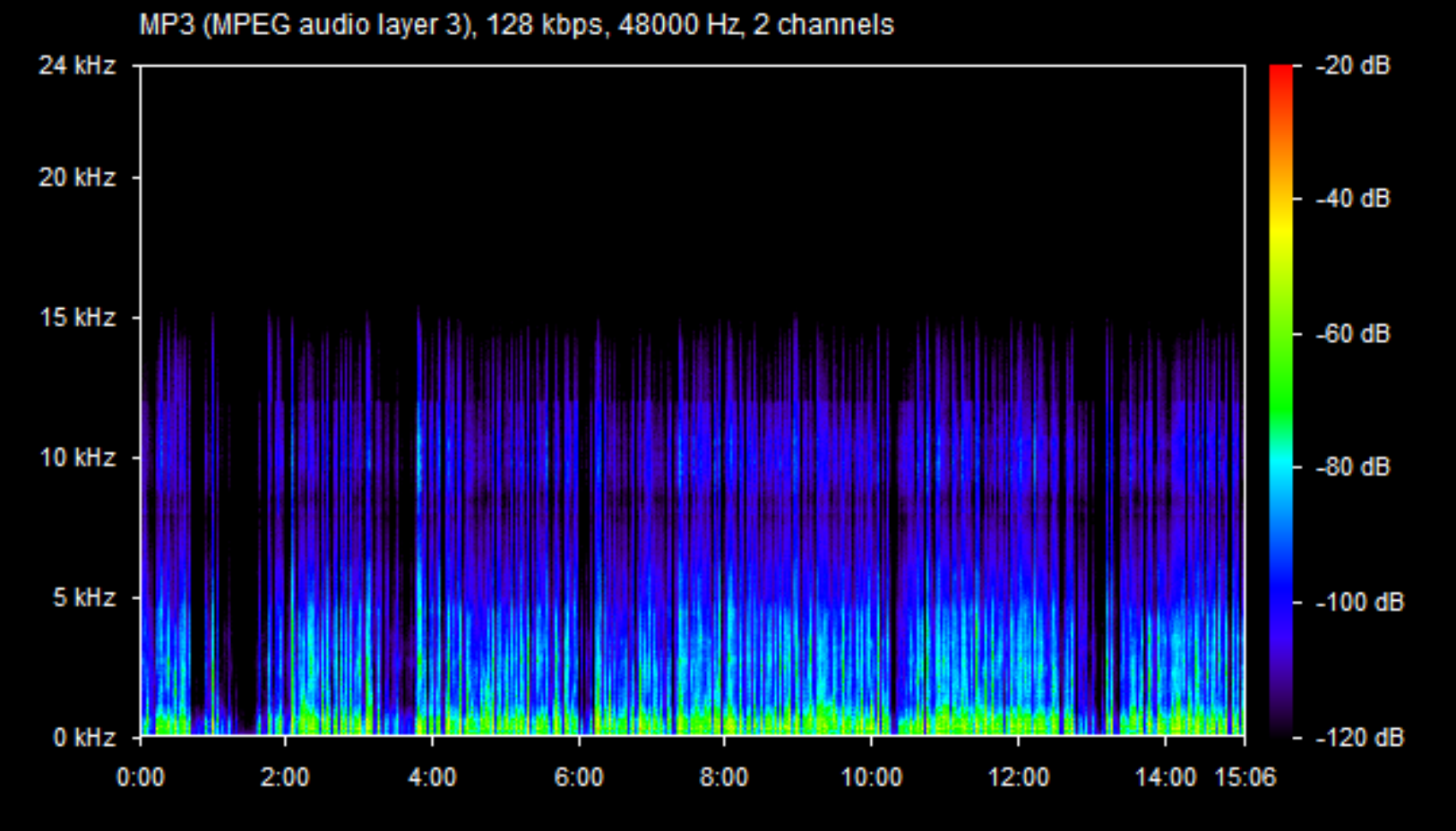}
    \caption{The original Audio Spectrum - The energy is distributed across various frequencies up to 24 KHz. The bitrate is 128 Kbps and the audio is dual channel.}\label{fig:spec_original}
\end{figure}
\begin{figure}
    \centering
    \includegraphics[scale=0.4]{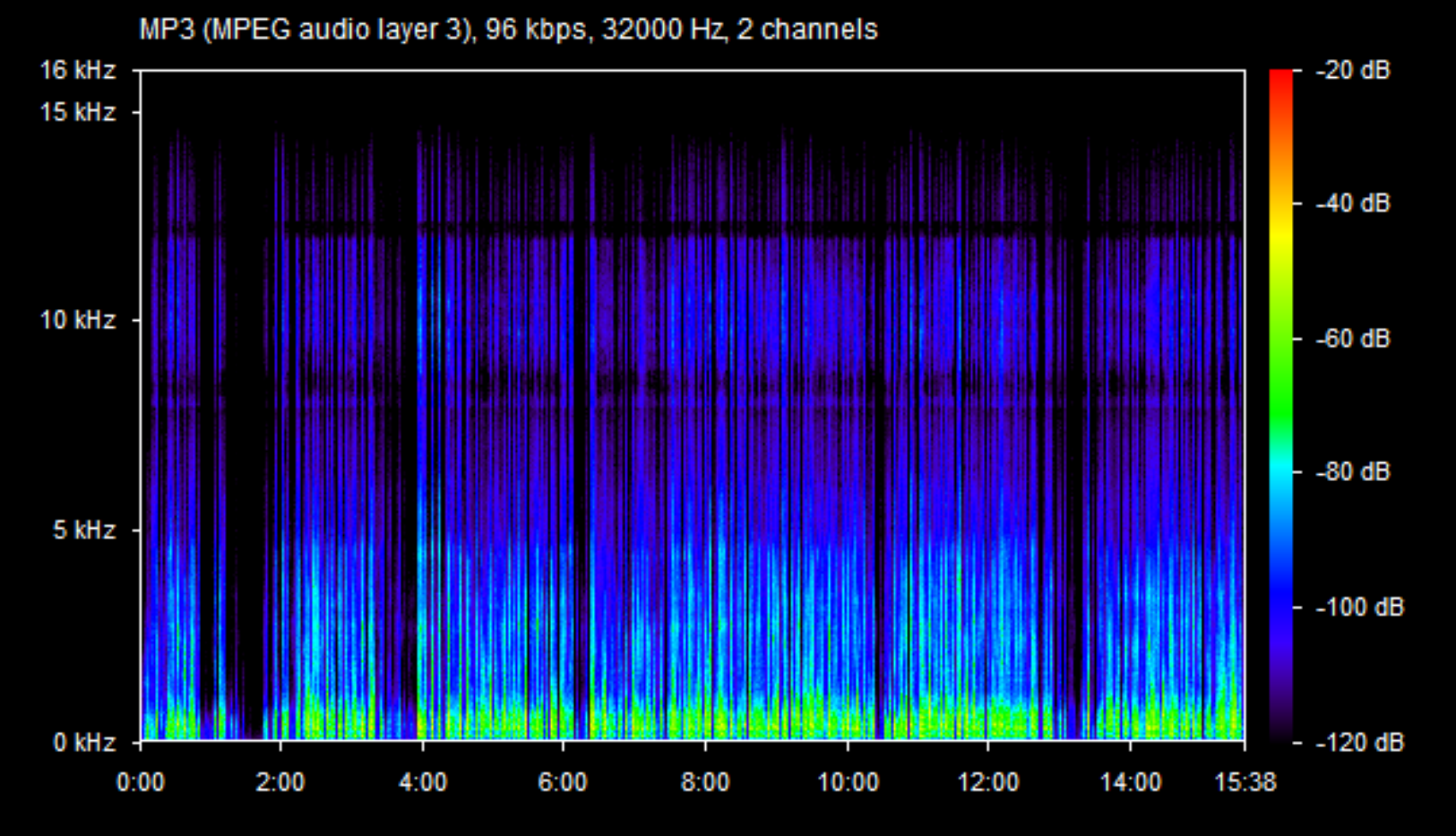}
    \caption{Google Meet Audio Spectrum over WiFi- The energy is distributed across various frequencies up to 16 KHz. The bitrate is 96 Kbps and the audio is dual channel. Compared to the original audio spectrum, the energy distribution is very similar.}\label{fig:spec_meet}
\end{figure}
\begin{figure}
    \centering
    \includegraphics[scale=0.4]{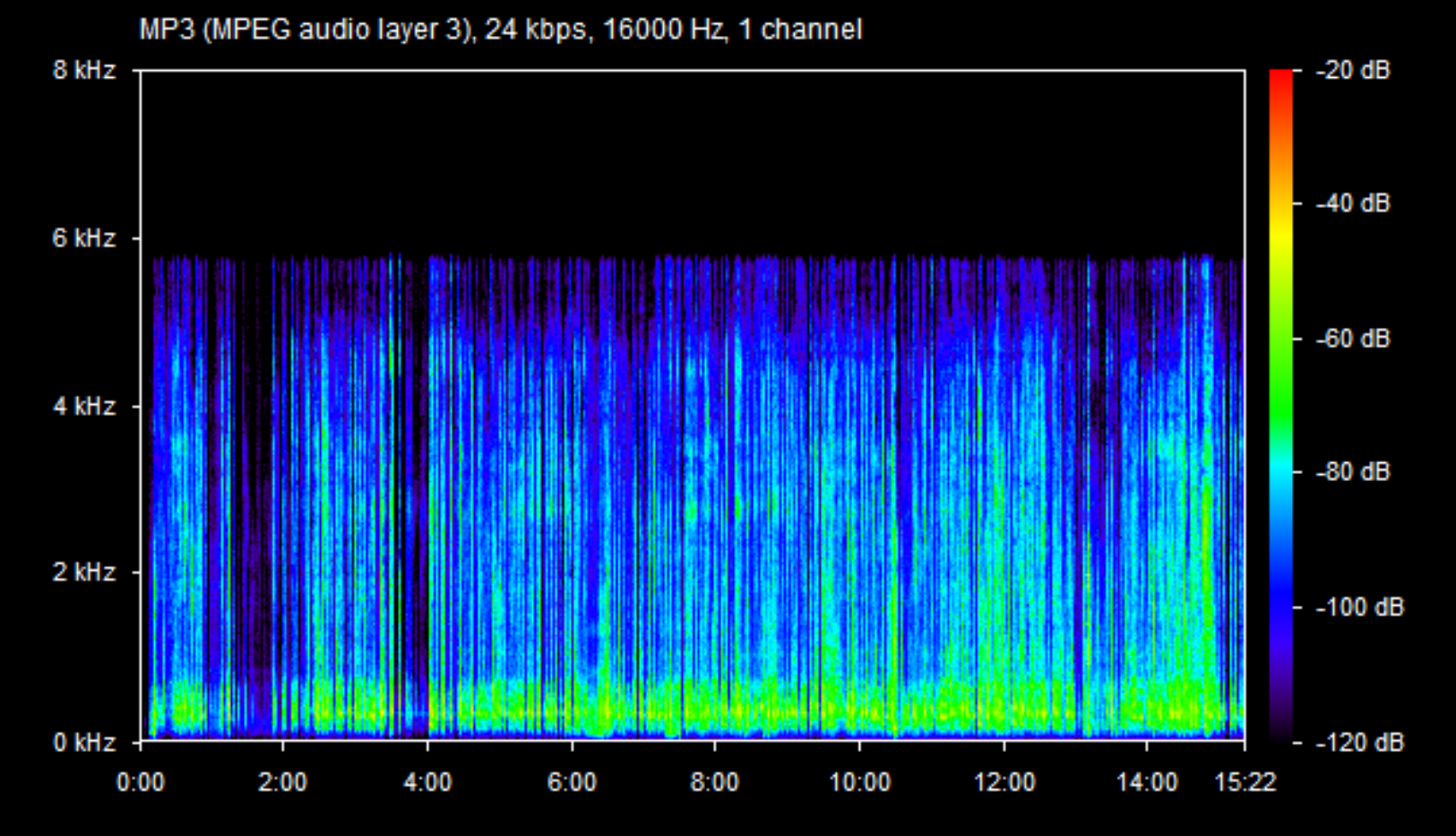}
    \caption{Microsoft Teams Audio Spectrum over WiFi- The energy is distributed across lesser frequencies up to 8 KHz. The bitrate is 24 Kbps and the audio is mono channel. Compared to the original audio spectrum, much energy is lost as most of the high frequency notes have been chopped off.}\label{fig:spec_teams}
\end{figure}
\begin{figure}
    \centering
    \includegraphics[scale=0.4]{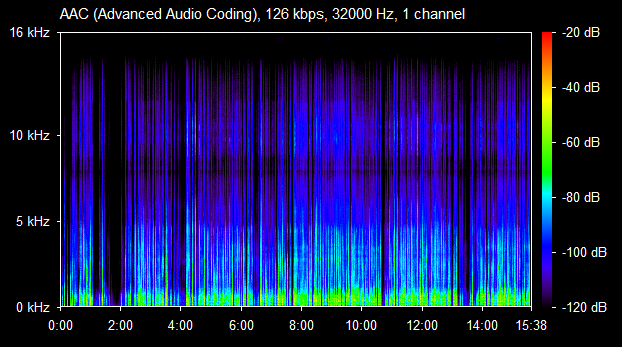}
    \caption{Zoom Audio Spectrum over WiFi- The energy is distributed across various frequencies up to 16 KHz. The bitrate is 126 Kbps and the audio is mono channel.}\label{fig:spec_zoom}
\end{figure}
\begin{figure}
    \centering
    \includegraphics[scale=0.5]{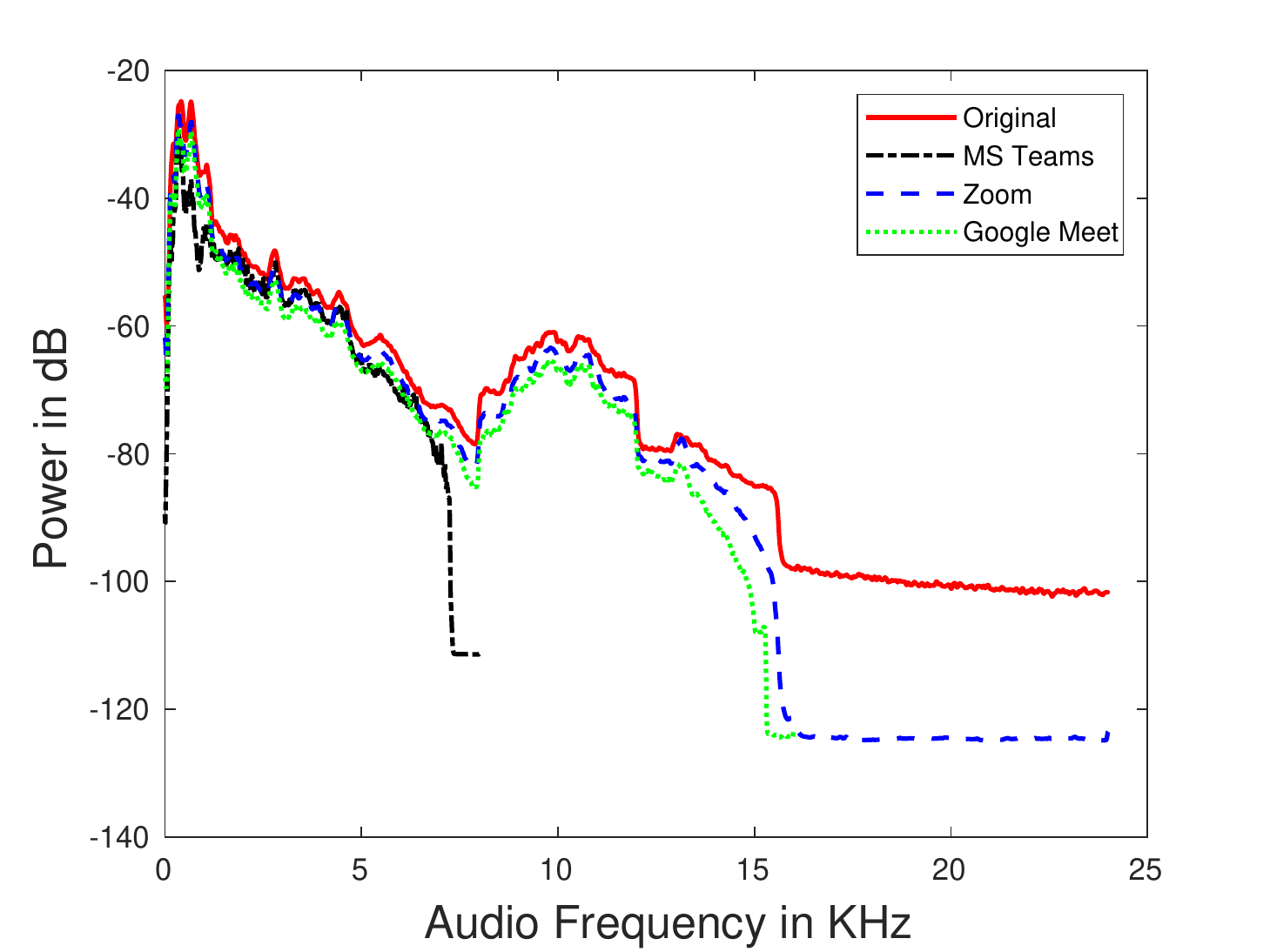}
    \caption{Microsoft Teams has a similar power level as compared to the original audio for the lower frequencies, but truncates all the higher frequencies above ~8 KHz. Zoom and Google Meet on the other hand retains some of the higher frequencies till about 16 KHz.}\label{fig:wifi_power}
\end{figure}

In terms of the audio quality, we find that audio content received over Google Meet and Zoom was much closer to its original content
, as observed in the spectrum graphs plotted in Figures \ref{fig:spec_original},\ref{fig:spec_meet}, \ref{fig:spec_teams}, and \ref{fig:spec_zoom}. 
On the other hand, Microsoft Teams compresses the audio to a greater extent by truncating the higher frequency notes. It can be seen in Figure \ref{fig:wifi_power} that Microsoft Teams has a higher power distribution amongst the frequencies below 8 KHz but does not have any higher frequencies after that. On the other hand, Google Meet and Zoom can produce frequencies closer to the original audio and truncate above 16 KHz, closer to human perceived frequencies of 20 KHz. The bitrate of Microsoft Teams was substantially lower than that of Google Meet and Zoom. Zoom and Microsoft Teams also down-converts the audio from a stereo channel to a mono channel. Upon performing a measurement involving only the mic with no screen video, Google Meet sent 6.42 MB of data from the sender to the server and 6.38 MB of data from the server to the receiver. Microsoft Teams sent 6.58 MB of data from the sender to the server but only 5.89 MB from the server to the receiver. Zoom sent 6.44 MB of data from the sender to the server and 6.36 MB from the server to the receiver. This indicates that Microsoft Teams performs higher compression at the server and then forwards the compressed packets to the receiver, lowering audio quality.  

\subsubsection{Utilization of Hardware Resources}
\begin{table*}
\centering
\newcolumntype{L}{>{\centering\arraybackslash}m{2.3cm}}
\caption{Sender-side resource consumption for wired broadband over WiFi. Microsoft Teams has a consistently high CPU utilization compared to Zoom and Google Meet. There is a significant increase in memory utilized by Google Meet when camera is switched ON.}
\label{tab:hardware_sender}
\begin{tabular}{cccLL}
\toprule
\textit{Test Type} & \textit{Platform} & \textit{CPU Load (\%)} & \textit{Memory Consumption (MB)} & \textit{Battery Consumption (\%)} \\
\midrule
\multirow{3}{*}{Mic OFF Cam OFF} & Google Meet & 13.35 & 336.61 & 5.00\\
 & MS Teams & 26.58 & 294.68 & 8.00\\
 & Zoom & 15.91 & 355.23 & 6.00\\
\hline
\multirow{3}{*}{Mic ON Cam OFF} & Google Meet & 22.68 & 388.71 & 6.00\\
 & MS Teams & 29.21 & 312.39 & 9.00\\
 & Zoom & 16.05 & 358.82 & 10.00\\
\hline
\multirow{3}{*}{Mic OFF Cam ON} & Google Meet & 25.14 & 555.40 & 7.00\\
 & Teams & 30.89 & 360.77 & 10.00\\
 & Zoom & 20.54 & 370.17 & 9.00\\
\hline
\multirow{3}{*}{Mic ON Cam ON} & Google Meet & 25.22 & 575.90 & 8.00\\
 & MS Teams & 31.74 & 372.20 & 10.00\\
 & Zoom & 22.88 & 382.85 & 10.00\\
\bottomrule
\end{tabular}
\end{table*}

\begin{table*}
\centering
\newcolumntype{L}{>{\centering\arraybackslash}m{2.3cm}}
\caption{Receiver-side resource consumption for wired broadband over WiFi. Microsoft Teams has low CPU utilization compared to Zoom and Google Meet. There is a significant increase in memory utilized by Google Meet when camera is switched ON.}
\label{tab:hardware_receiver}
\begin{tabular}{cccLL}
\toprule
\textit{Test Type} & \textit{Platform} & \textit{CPU Load (\%)} & \textit{Memory Consumption (MB)} & \textit{Battery Consumption (\%)} \\
\midrule
\multirow{3}{*}{Mic OFF Cam OFF} & Google Meet & 2.31 & 233.13 & 5.00\\
 & MS Teams & 4.64 & 209.33 & 7.00\\
 & Zoom & 9.93 & 232.85 & 6.00\\
\hline
\multirow{3}{*}{Mic ON Cam OFF} & Google Meet & 6.07 & 269.09 & 6.00\\
 & MS Teams & 6.90 & 222.17 & 7.00\\
 & Zoom & 11.52 & 246.97 & 7.00\\
\hline
\multirow{3}{*}{Mic OFF Cam ON} & Meet & 12.99 & 489.22 & 7.00\\
 & Teams & 9.11 & 275.53 & 8.00\\
 & Zoom & 13.81 & 258.24 & 8.00\\
\hline
\multirow{3}{*}{Mic ON Cam ON} & Google Meet & 13.51 & 514.17 & 7.00\\
 & MS Teams & 9.98 & 291.55 & 8.00\\
 & Zoom & 14.30 & 263.09 & 8.00\\
\bottomrule
\end{tabular}
\end{table*}

\begin{figure}
    \centering
    \includegraphics[scale=0.4]{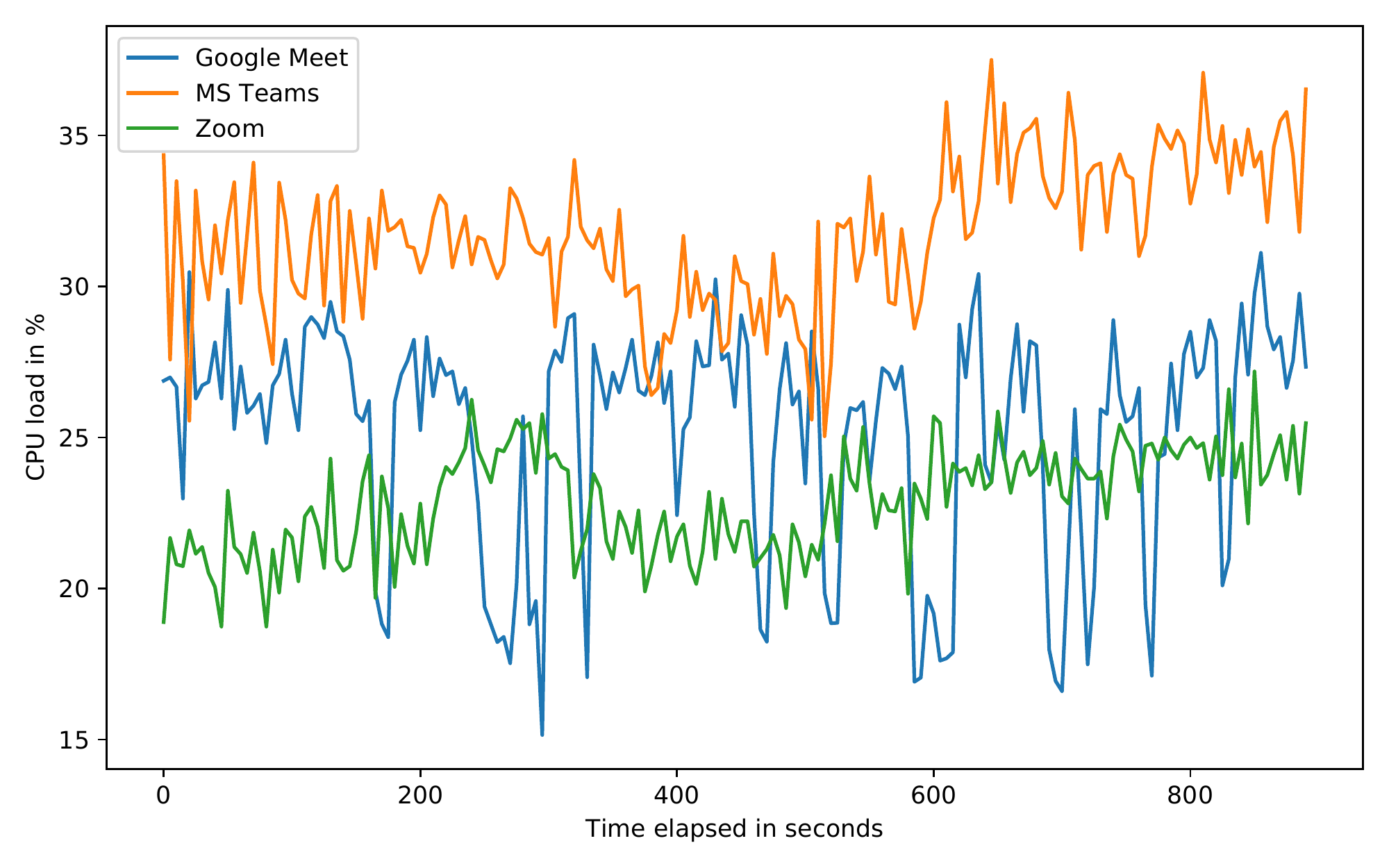}
    \caption{Sender-side CPU load for WiFi with mic ON and camera ON. Microsoft Teams uses the highest amount of CPU resources on the sender-side. Google Meet and Zoom have a similar average CPU utilization, however, CPU load has lesser variation for Zoom. Occasional spikes and dips can be seen for Google Meet. }\label{fig:cpu_sender}
\end{figure}
\begin{figure}
    \centering
    \includegraphics[scale=0.4]{ 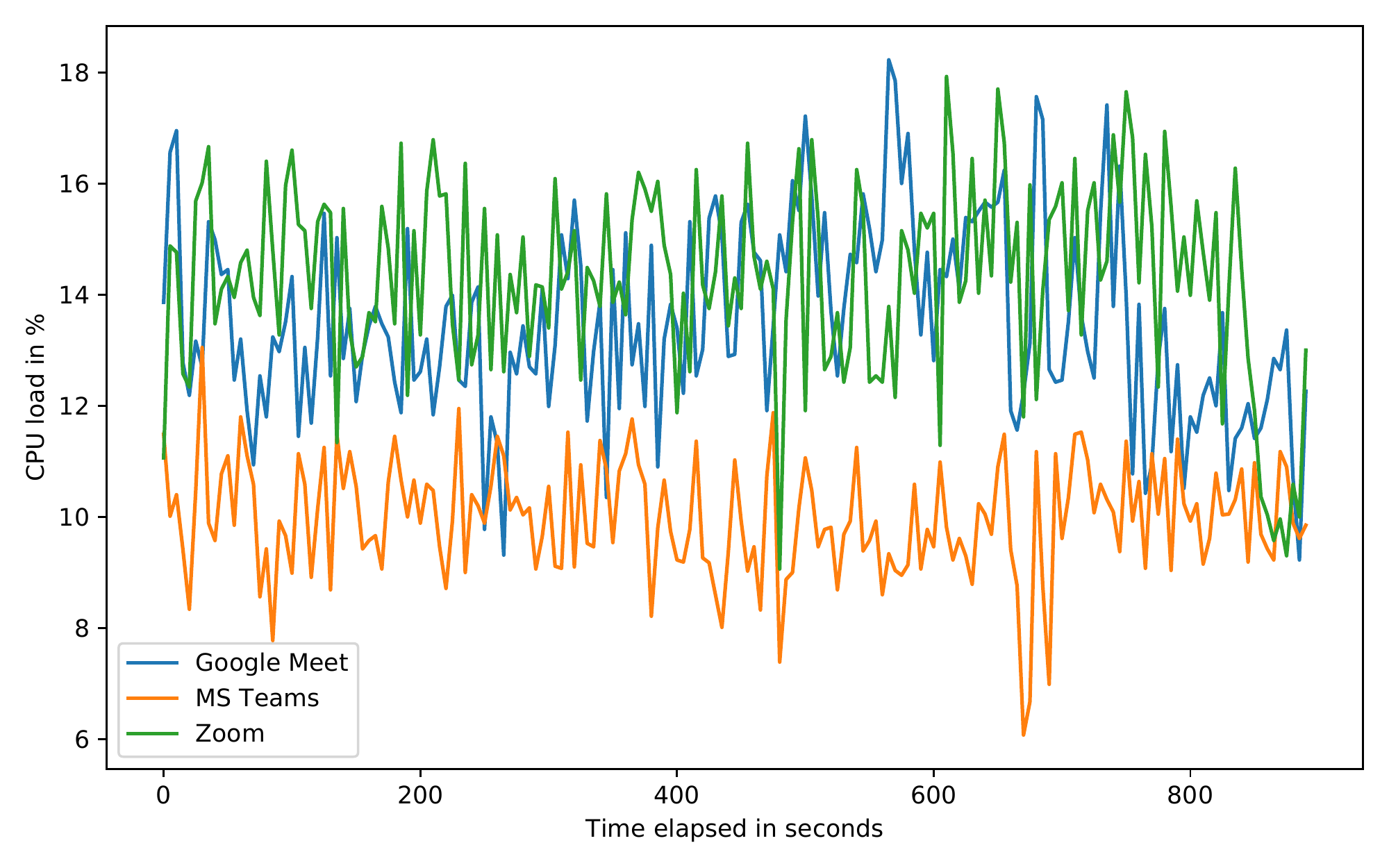}
    \caption{Receiver-side CPU load for WiFi with mic ON and cam ON. Microsoft Teams uses the least amount of CPU resources on the receiver-side. Google Meet and Zoom have a similar CPU utilization.}\label{fig:cpu_receiver}
\end{figure}

We summarise the quantitative measurements of the resource consumption by the apps over the wired broadband network with end-host connected by WiFi in Tables \ref{tab:hardware_sender} and \ref{tab:hardware_receiver}. 
On the sender-side, in addition to sending high payload, Microsoft Teams has the highest CPU load among all the apps for all test types.
Figure \ref{fig:cpu_sender} plots the CPU utilization for the sender when the mic and camera are switched ON. Google Meet's memory consumption increases more than any other app when the camera is turned ON, as it compresses the video aggressively. 
Zoom has a consistent memory usage even with an increase in payload. However, it consumes more memory than the other two apps when the mic and camera are switched off. The battery consumption increases for all the apps when the camera is turned ON.

On the receiver side, we see that Zoom has the highest CPU usage, while Microsoft Teams has the least as it receives audio with lesser content. Figure \ref{fig:cpu_receiver} plots the CPU utilization for the receiver when the mic and camera are switched ON. There is a general trend of slightly lower battery consumption by all three apps compared to sender, especially in the case of Zoom and MS Teams. Zoom and Microsoft Teams also use nearly the same amount of memory. However, Google Meet has a highly variable memory usage depending on payload, with an increase of 75\% when the camera is switched ON. CPU, memory, and battery usage trend upwards with increasing payload for all three apps.


\subsection{Quantitative Performance over 4G Mobile Internet}

\begin{table*}
\centering
\caption{Summary of data collected over 4G mobile Internet. Google Meet significantly reduces the payload sizes whereas Microsoft Teams and Zoom continue to send similar payload as that for wired broadband. Microsoft Teams uses higher data bandwidth than Google Meet and Zoom. Zoom has a consistently higher PSNR and SSIM than Google Meet and Microsoft Teams. Unlike the case of Wired Broadband over WiFi, Microsoft Teams has the lowest standard deviation in IPAT.}\label{tab:results_4g}
\begin{tabular}{ccccccc}
\toprule
Measurement Type & App & \begin{tabular}[c]{@{}l@{}}Download\\ Payload (MB)\end{tabular} & \begin{tabular}[c]{@{}l@{}}Upload\\ Payload (MB)\end{tabular} & \begin{tabular}[c]{@{}l@{}}$\sigma({IPAT})$\\ (ms)\end{tabular} & \begin{tabular}[c]{@{}l@{}}PSNR\\(YUV)\end{tabular} & 
\begin{tabular}[c]{@{}l@{}}SSIM\\(YUV)\end{tabular} \\
\midrule
\multirow{2}{*}{Mic OFF Cam OFF} & Google Meet & 41 & 48 & 34.75 & 28.50 & 0.88 \\
 & MS Teams & 71 & 104 & 30.76 & 35.96 & 0.99 \\
 & Zoom & 53 & 57 & 34.2 & 49.44 & 0.99\\
\hline
\multirow{2}{*}{Mic ON Cam OFF} & Google Meet & 43 & 50 & 31.59 & 27.96 & 0.86 \\
 & MS Teams & 82 & 112 & 12.95 & 33.85 & 0.98 \\
 & Zoom & 58 & 61 & 23.3 & 48.47 & 0.99\\
\hline
\multirow{2}{*}{Mic OFF Cam ON} & Google Meet & 46 & 57 & 25.22 & 25.17 & 0.83 \\
 & MS Teams & 96 & 119 & 11.95 & 30.74 & 0.95 \\
 & Zoom & 75 & 82 & 19.1 & 47.85 & 0.99\\
\hline
\multirow{2}{*}{Mic ON Cam ON} & Google Meet & 49 & 59 & 23.83 & 24.96 & 0.82 \\
 & MS Teams & 98 & 122 & 10.98 & 29.98 & 0.95 \\
 & Zoom & 78 & 84 & 18.8 & 47.53 & 0.99\\
\bottomrule
\end{tabular}
\end{table*}

\begin{table}
\centering
\caption{The values of Y, U, and V towards the computation of SSIM. We find a significant dip in SSIM 'Y' value for Google Meet when camera is switched ON.}
\label{tab:yuv_4g}
\begin{tabular}{ccccc}
\toprule
Test Type & Platform & Y & U & V \\
\midrule

\multirow{3}{*}{Mic OFF Cam OFF} & Google Meet & 0.83 & 0.98 & 0.98 \\
 & MS Teams & 0.98 & 0.99 & 0.99 \\
 & Zoom & 0.98 & 0.99 & 0.99\\
\hline
\multirow{3}{*}{Mic ON Cam OFF} & Google Meet & 0.82 & 0.99 & 0.99 \\
 & MS Teams & 0.97 & 0.99 & 0.99 \\
 & Zoom & 0.98 & 0.99 & 0.99\\
\hline
\multirow{3}{*}{Mic OFF Cam ON} & Meet & 0.61 & 0.99 & 0.97 \\
 & Teams & 0.95 & 0.99 & 0.98 \\
 & Zoom & 0.98 & 0.99 & 0.99 \\
\hline
\multirow{3}{*}{Mic ON Cam ON} & Google Meet & 0.60 & 0.98 & 0.97 \\
 & MS Teams & 0.95 & 0.99 & 0.99 \\
 & Zoom & 0.99 & 0.99 & 0.99 \\
\bottomrule
\end{tabular}
\end{table}

\begin{figure}
    \centering
    \includegraphics[scale=0.5]{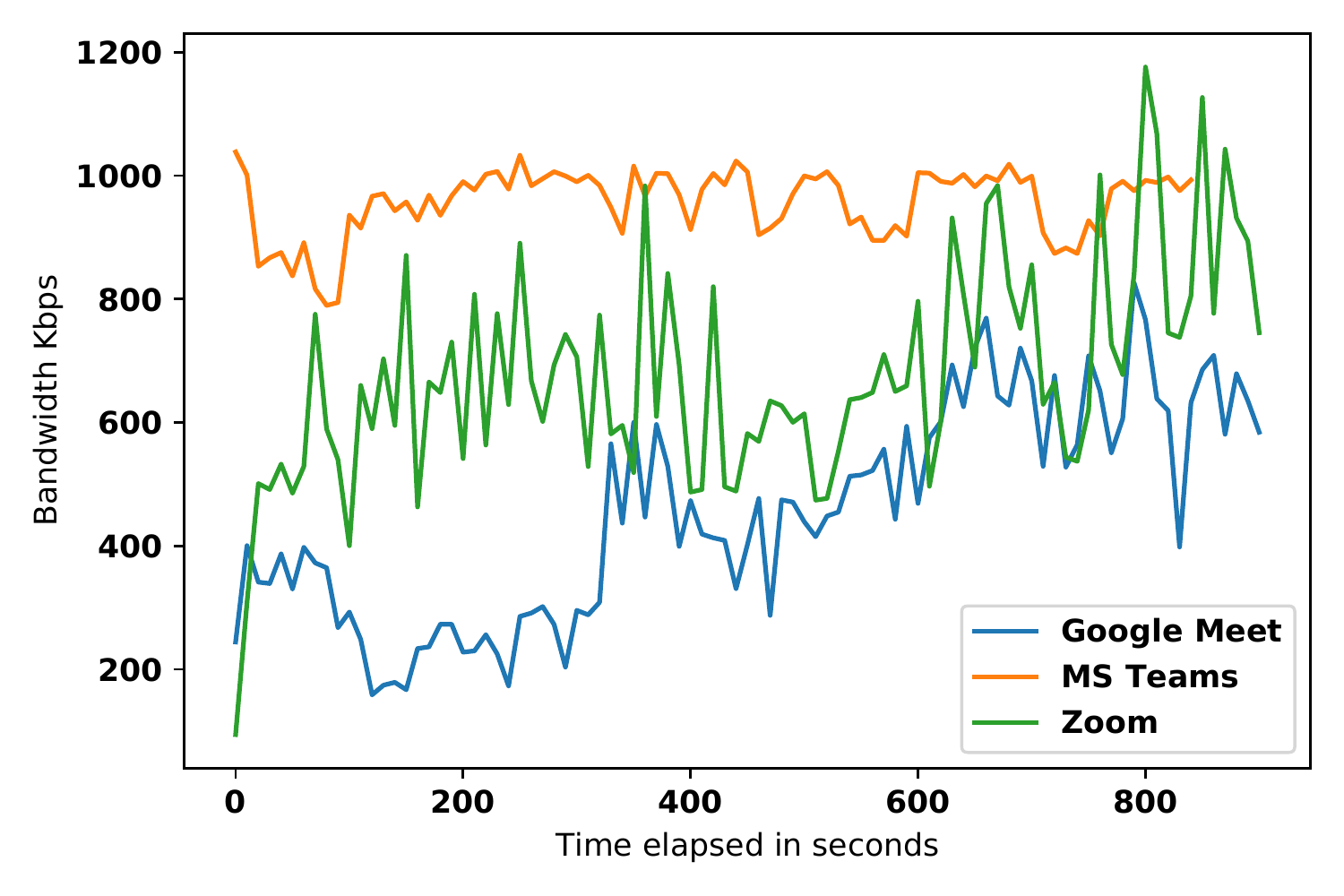}
    \caption{We plot Download Bandwidth every 10 sec for 4G mobile Internet. Microsoft Teams uses more bandwidth and is more stable than Google Meet and Zoom. Zoom uses more bandwidth than Google Meet but is also more unstable. The standard deviation in Google Meet bandwidth is  198.64 Kbps, Microsoft Teams is 105.52 Kbps, and Zoom is 222.76 Kbps.}\label{fig:dl_bw_4g}
\end{figure}
\begin{figure}
    \centering
    \includegraphics[scale=0.5]{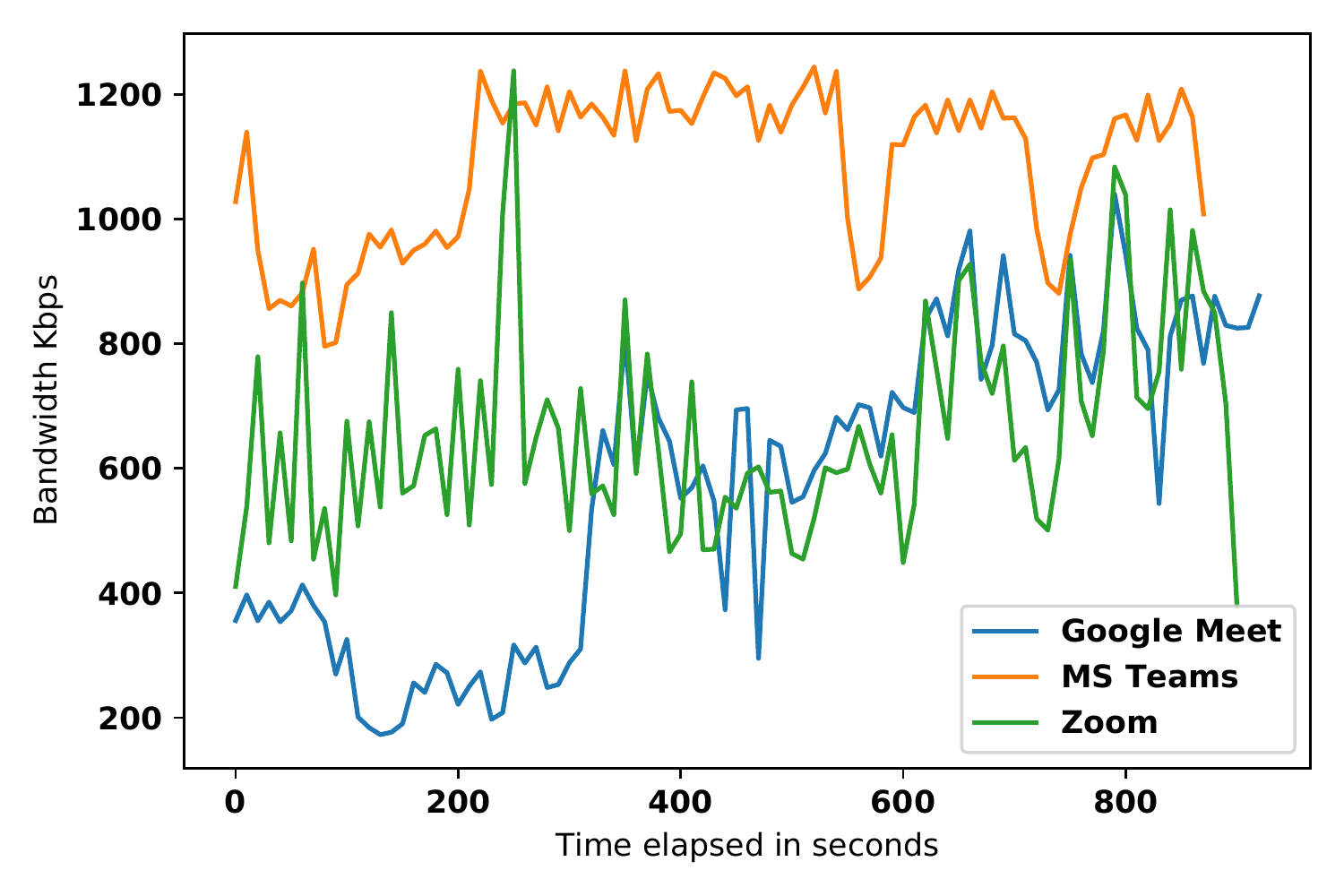}
    \caption{We plot the Upload Bandwidth every 10 sec for 4G mobile Internet. Microsoft Teams uses more bandwidth and is more stable than Google Meet and Zoom. Zoom uses more data than Google Meet. Zoom and Google Meet are almost equally stable. The standard deviation in Google Meet bandwidth is 229.16 Kbps, Microsoft Teams is 156.16 Kbps, and Zoom is 208.22 Kbps.}\label{up_bw_4g}
\end{figure}

We summarize the results from the measurements on 4G mobile Internet in Table \ref{tab:results_4g}. We observe that over a 4G mobile Internet connection, Microsoft Teams and Zoom perform nearly the same over wired broadband. However, there is a significant dip in Google Meet's performance. The bandwidth for Microsoft Teams is stable compared to that of Google Meet and Zoom 
as seen in Figures \ref{fig:dl_bw_4g} and \ref{up_bw_4g} when the mic and the camera are switched ON. We see similar plots for other measurements.

Google Meet reduces the Upload Payload, which drops the transmission rate by almost 40\%. This strategy is helpful considering that the 4G mobile Internet is more expensive than wired broadband. 
However, due to these optimizations, the video quality in Google Meet is reduced since it was optimizing heavily on the luminescence of the video, as shown by the SSIM Y values
in Table \ref{tab:yuv_4g}. 
In contrast, Microsoft Teams and Zoom perform better with almost no change in the video quality between wired broadband and 4G Internet. In addition, the standard deviation in IPAT is less for Microsoft Teams than Google Meet and Zoom, which affects the video quality. Zoom continues to have a better video quality with lower data utilization, as can be seen by the PSNR values. The standard deviation in IPAT almost doubled for Google Meet and Zoom when switching from wired broadband to 4G mobile Internet and remains comparatively higher for all the measurements. This suggests that Google Meet and Zoom do not transmit the packets at a steady rate for a 4G mobile Internet connection \cite{bouraqia-2020}.

\begin{figure}
    \centering
    \includegraphics[scale=0.4]{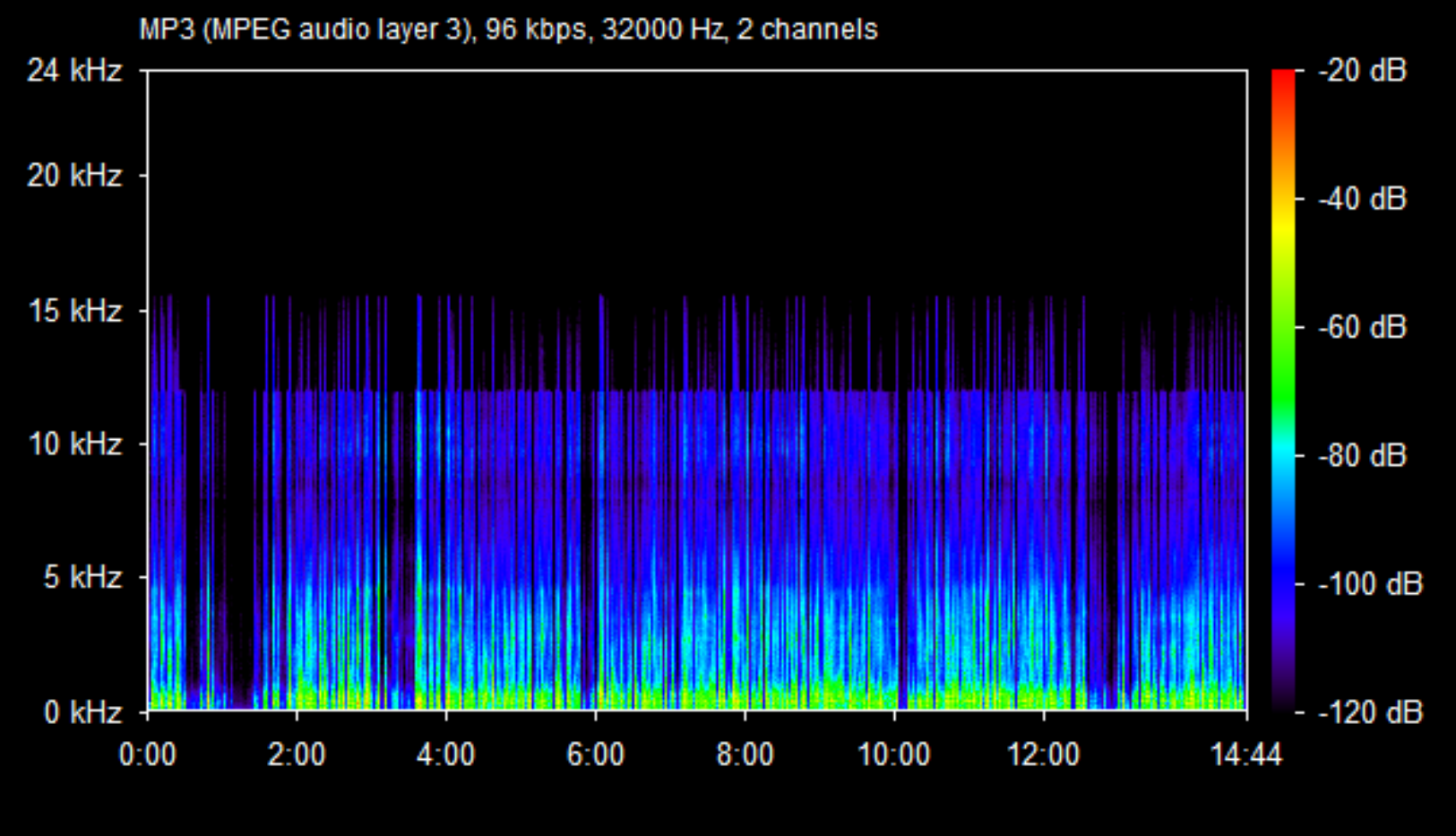}
    \caption{Google Meet Audio Spectum 4G- The energy is distributed across various frequencies up to 16 KHz. The bitrate is 96 Kbps and the audio is dual channel. Compared to the original audio spectrum, the energy distribution is very similar.}\label{fig:spec_meet4g}
\end{figure}
\begin{figure}
    \centering
    \includegraphics[scale=0.4]{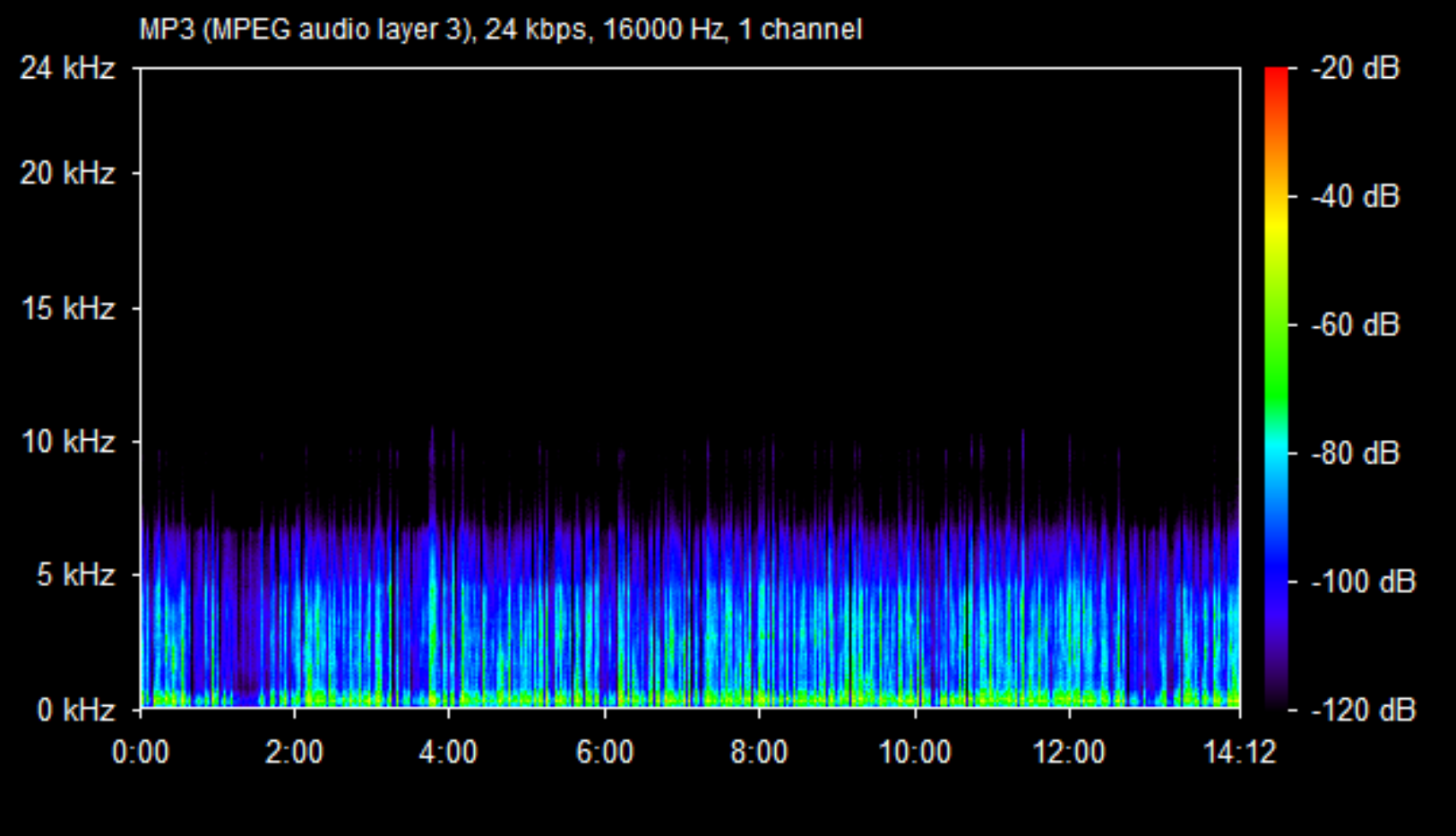}
    \caption{Microsoft Teams Audio Spectrum 4G- The energy is distributed across lesser frequencies up to 8 KHz. The bitrate is 24 Kbps and the audio is mono channel. Compared to the original audio spectrum, the energy lost is much higher since most of the high frequency notes have been chopped off. }\label{fig:spec_teams4g}
\end{figure}
\begin{figure}
    \centering
    \includegraphics[scale=0.4]{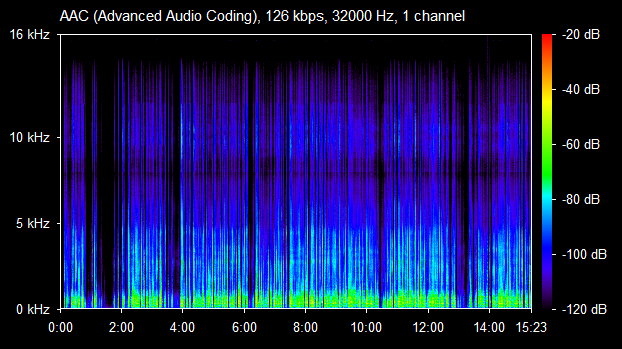}
    \caption{Zoom Audio Spectrum 4G - The energy is distributed across various frequencies up to 16 KHz. The bitrate is 126 Kbps and the audio is mono channel.}\label{fig:spec_zoom4g}
\end{figure}
\begin{figure}
    \centering
    \includegraphics[scale=0.5]{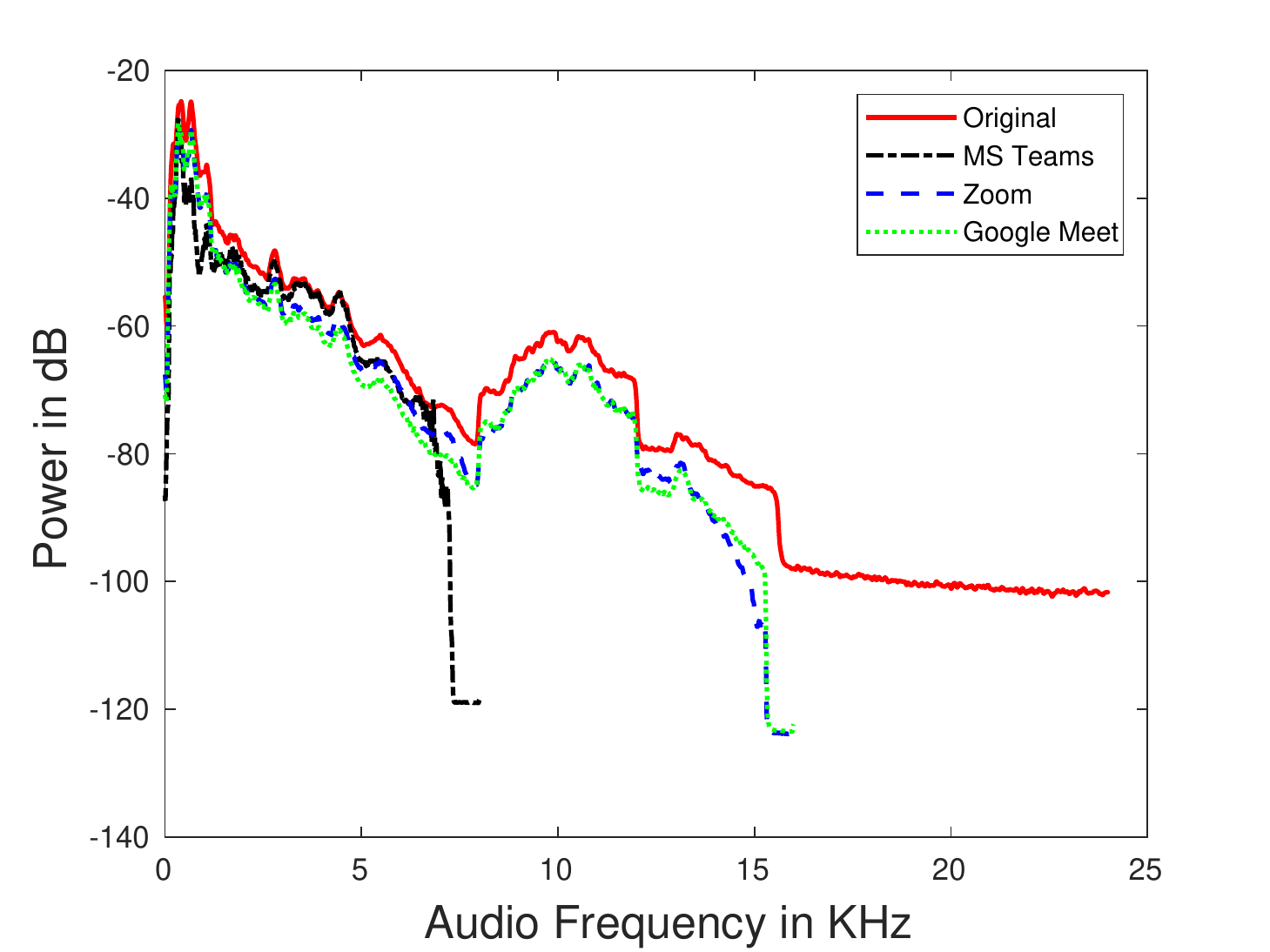}
    \caption{Microsoft Teams has a similar power level as compared to the original audio for the lower frequencies, but truncates all the higher frequencies above ~8 KHz. The same behavior was seen for the case of wired broadband over WiFi. Google Meet and Zoom retain some of the higher frequencies till about 16 KHz, however, the power for the higher frequencies is slightly lower than the original power.}\label{fig:4g_power}
\end{figure}

We show the audio spectrum graphs for Google Meet, Microsoft Teams, and Zoom over 4G mobile Internet connections in Figures \ref{fig:spec_meet4g}, \ref{fig:spec_teams4g}, and \ref{fig:spec_zoom4g}. 
In terms of audio quality, the observations are similar to the case of wired broadband over WiFi. Microsoft Teams again chopped off the higher frequency notes and has a more concise energy distribution. On the contrary, Google Meet and Zoom keep the audio quality closer to the original with a more dispersed energy distribution, as seen in Figure \ref{fig:4g_power}. However, Google Meet and Zoom dropped the power of frequencies above 10 KHz by approximately 5\% compared to broadband. The bitrate for Microsoft Teams continues to be lower than that for Google Meet and Zoom. Microsoft Teams and Zoom down-convert the audio from a stereo channel to a mono channel. Upon performing a measurement involving only mic and no screen sharing, we see that Google Meet sent 5.79 MB of data from the sender to the server and 5.74 MB of data from the server to the receiver. Zoom sent 5.91 MB of data from the sender to the server and 5.88 MB from the server to the receiver. In comparison, Microsoft Teams sent 6.78 MB of data from the sender to the server and 6.06 MB of data from the server to the receiver. Like in the case of wired broadband over WiFi, Microsoft Teams is performing compression at the server and then forwarding the packets to the receiver, resulting in poor audio quality.




\subsubsection{Utilization of Hardware Resources}

\begin{table*}
\centering
\newcolumntype{L}{>{\centering\arraybackslash}m{2.3cm}}
\caption{Sender-side resource consumption for 4G mobile Internet. Google Meet and Zoom have a similar CPU usage, whereas Microsoft Teams has a comparatively higher CPU usage. It has a low memory consumption. Google Meet's memory consumption increases significantly when the camera is switched ON.}
\label{tab:hardware_sender_4g}
\begin{tabular}{cccLL}
\toprule
\textit{Test Type} & \textit{Platform} & \textit{CPU Load (\%)} & \textit{Memory Consumption (MB)} & \textit{Battery Consumption (\%)} \\
\midrule
\multirow{3}{*}{Mic OFF Cam OFF} & Google Meet & 12.457 & 336.688 & 7.00\\
 & MS Teams & 20.171 & 243.624 & 8.00\\
 & Zoom & 10.109 & 276.107 & 7.00\\
\hline
\multirow{3}{*}{Mic ON Cam OFF} & Google Meet & 12.820 & 356.454 & 8.00\\
 & MS Teams & 26.993 & 267.860 & 9.00\\
  & Zoom & 12.861 & 296.275 & 8.00\\
\hline
\multirow{3}{*}{Mic OFF Cam ON} & Google Meet & 19.182 & 455.40 & 7.00\\
 & MS Teams & 31.294 & 365.27 & 9.00\\
 & Zoom & 17.327 & 452.39 & 10.00\\
\hline
\multirow{3}{*}{Mic ON Cam ON} & Google Meet & 19.692 & 517.890 & 9.00\\
 & MS Teams & 33.321 & 394.626 & 10.00\\
 & Zoom & 18.826 & 460.348 & 11.00\\
\bottomrule
\end{tabular}
\end{table*}

\begin{table*}
\centering
\newcolumntype{L}{>{\centering\arraybackslash}m{2.3cm}}
\caption{Receiver-side resource consumption for 4G mobile Internet. All apps have a similar CPU usage with Microsoft Teams having slightly lower than Google Meet and Zoom. Google Meet has a comparatively higher memory consumption. Google Meet's memory consumption increases significantly when the camera is switched on.}
\label{tab:hardware_receiver_4g}
\begin{tabular}{cccLL}
\toprule
\textit{Test Type} & \textit{Platform} & \textit{CPU Load (\%)} & \textit{Memory Consumption (MB)} & \textit{Battery Consumption (\%)} \\
\midrule
\multirow{3}{*}{Mic OFF Cam OFF} & Google Meet & 5.708 & 224.166 & 6.00\\
 & Teams & 4.920 & 211.205 & 7.00\\
 & Zoom & 6.496 & 222.839 & 6.00\\
\hline
\multirow{3}{*}{Mic ON Cam OFF} & Google Meet & 6.227 & 228.776 & 6.00\\
 & MS Teams & 5.938 & 217.568 & 8.00\\
 & Zoom & 6.949 & 246.366 & 7.00\\
\hline
\multirow{3}{*}{Mic OFF Cam ON} & Google Meet & 9.221 & 439.14 & 8.00\\
 & MS Teams & 7.784 & 291.253 & 8.00\\
 & Zoom & 7.454 & 248.44 & 8.00\\
\hline
\multirow{3}{*}{Mic ON Cam ON} & Google Meet & 10.131 & 466.703 & 9.00\\
 & MS Teams & 8.129 & 293.862 & 9.00\\
 & Zoom & 10.306 & 261.956 & 9.00\\
\bottomrule
\end{tabular}
\end{table*}

\begin{figure}
    \centering
    \includegraphics[scale=0.4]{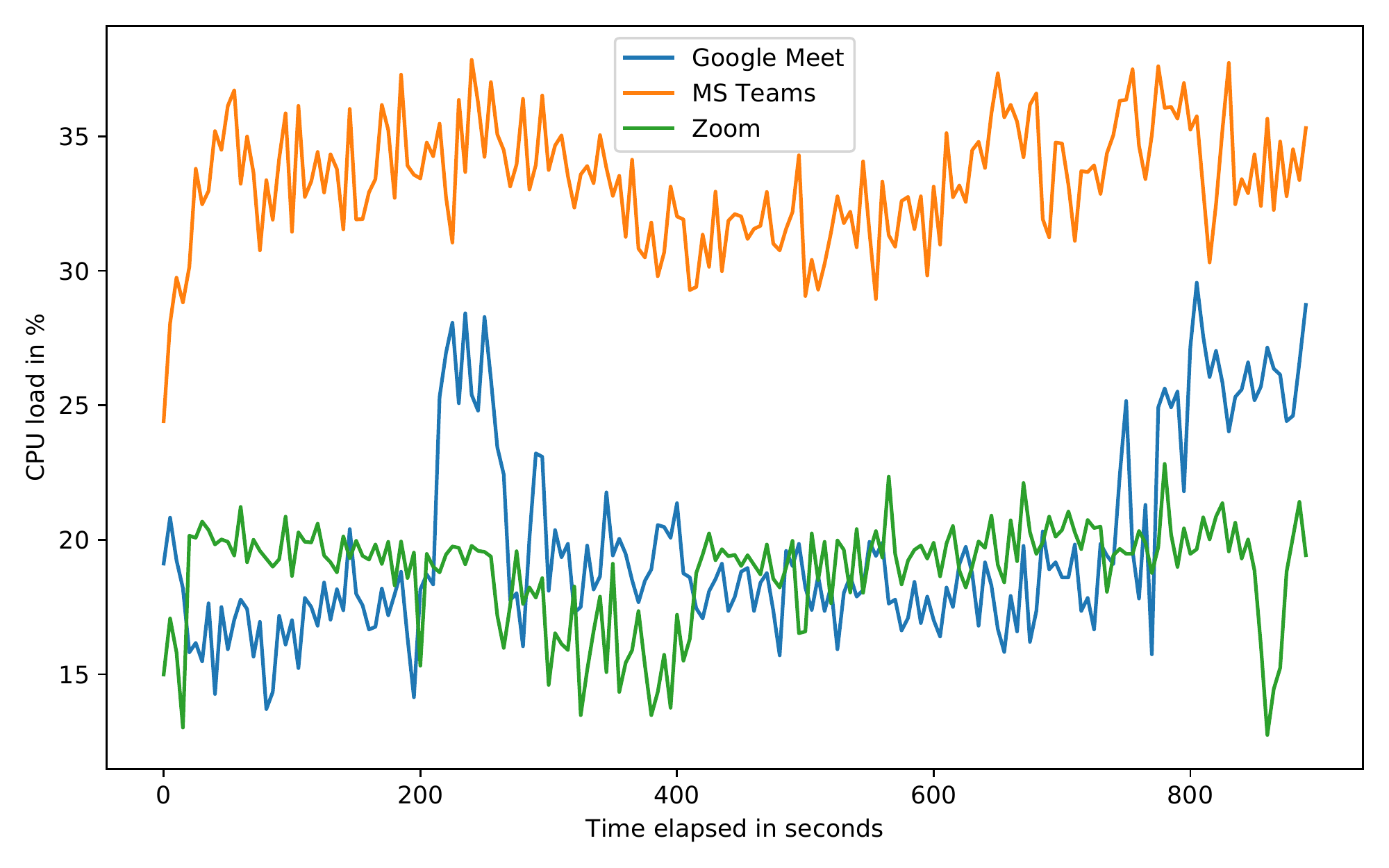}
    \caption{Sender-side CPU utilisation for 4G mobile Internet. Microsoft Teams has the maximum amount of CPU resources. Zoom and Google Meet consumed nearly the same amount of CPU, nearly half of that of Microsoft Teams.}\label{fig:cpu_sender_4g}
\end{figure}
\begin{figure}
    \centering
    \includegraphics[scale=0.4]{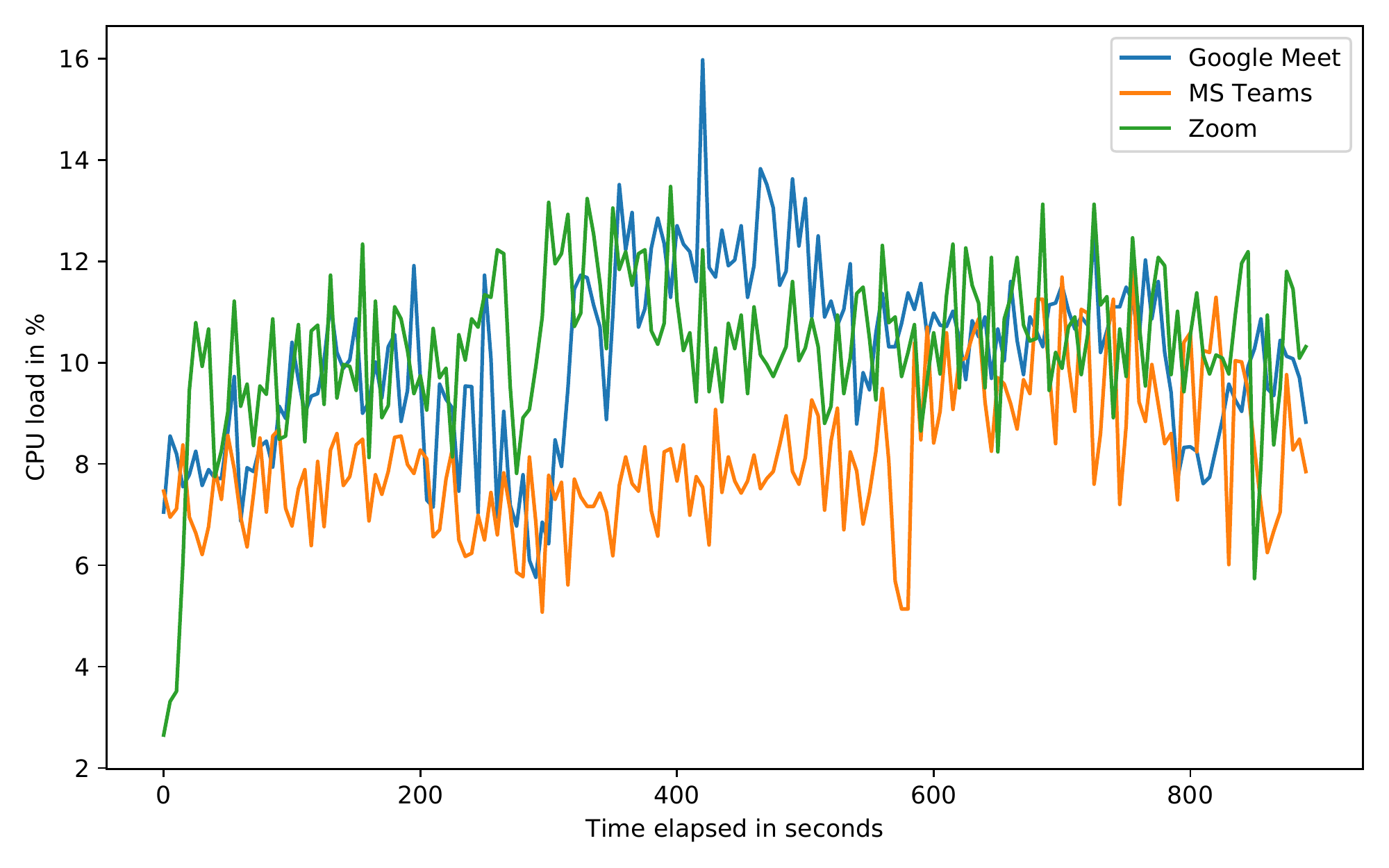}
    \caption{Receiver-side CPU utilisation for 4G mobile Internet. Microsoft Teams consumes the least amount of CPU resources. Zoom and Google Meet consumed nearly the same amount of CPU. Overall, there was not much difference for CPU utilization by all three applications.}\label{fig:cpu_receiver_4g}
\end{figure}

We summarise the quantitative measurements of the resource consumption by the apps over 4G mobile Internet in Tables \ref{tab:hardware_sender_4g} and \ref{tab:hardware_receiver_4g}. Microsoft Teams has a similar resource consumption on the sender-side as that for wired broadband over WiFi as it has a similar heavy payload.
However, Google Meet and Zoom show a decline in CPU usage. The fall is most significant when the camera is switched ON. This indicates a dip in video rendering as 4G video quality is lesser than broadband. Figure \ref{fig:cpu_sender_4g} plots the CPU utilization for the sender when the mic and camera are switched ON. Google Meet shows a slight decrease in memory consumption over 4G, while Zoom shows a slight increase.

On the receiver side, Microsoft Teams again has a similar usage as that for wired broadband over WiFi, whereas Google Meet and Zoom showed a decline in CPU usage. Figure \ref{fig:cpu_receiver_4g} plots the CPU utilization for the receiver when the mic and camera are switched ON. Google Meet also shows a significant dip in memory consumption. However, Zoom uses nearly the same amount of memory. 

\subsection{Qualitative Performance over Wired Broadband via WiFi and Mobile 4G Internet}

\begin{figure}
    \centering
    \includegraphics[scale=0.165]{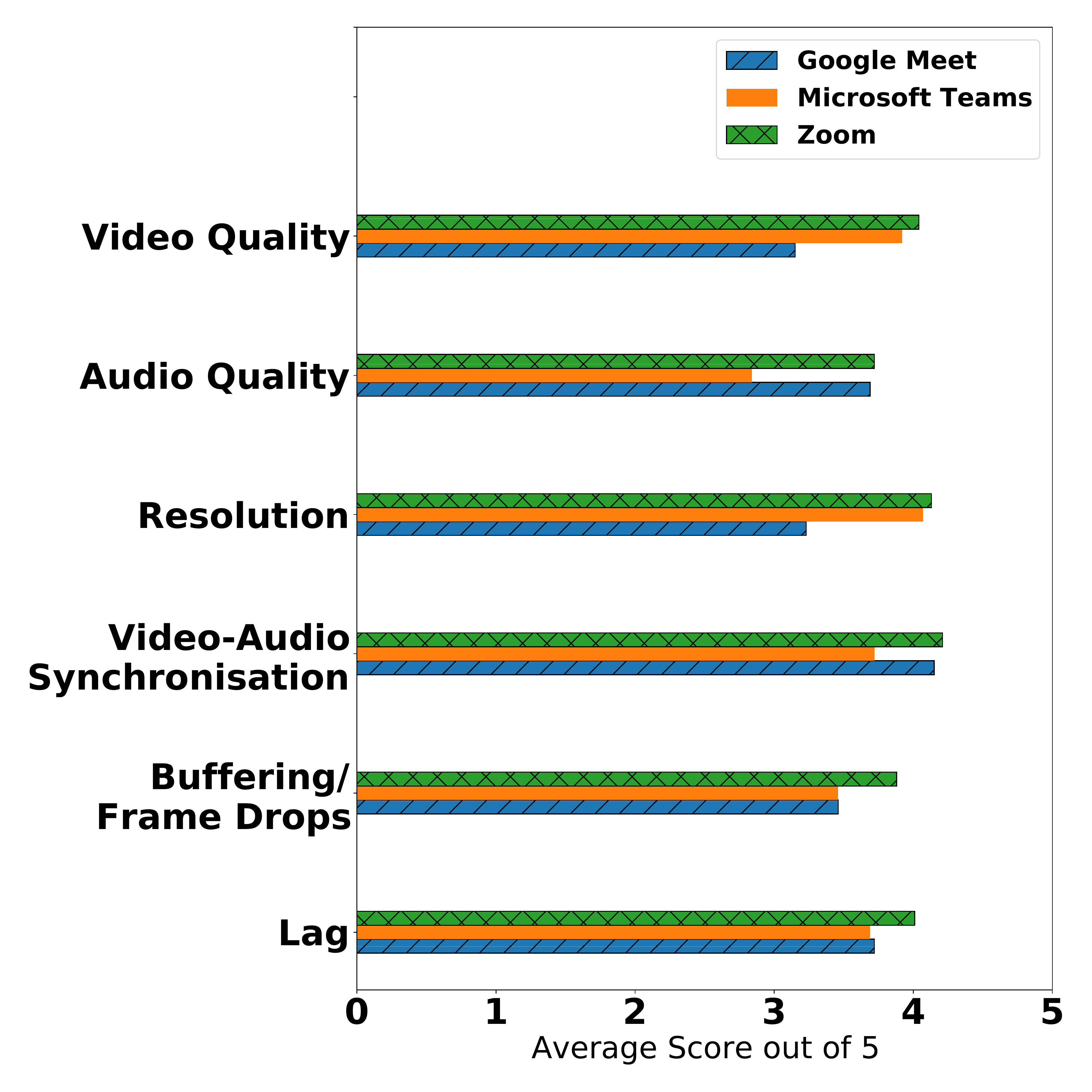}
    \caption{Qualitative Analysis Responses for Wired Broadband. The figure shows average score out of 5. Zoom provides a better video quality and resolution, Microsoft Teams is the second choice of users for video quality. Google Meet and Zoom have a better audio quality and video-audio synchronisation.}\label{fig:subjective_bb}
\end{figure}

\begin{figure}
    \centering
    \includegraphics[scale=0.165]{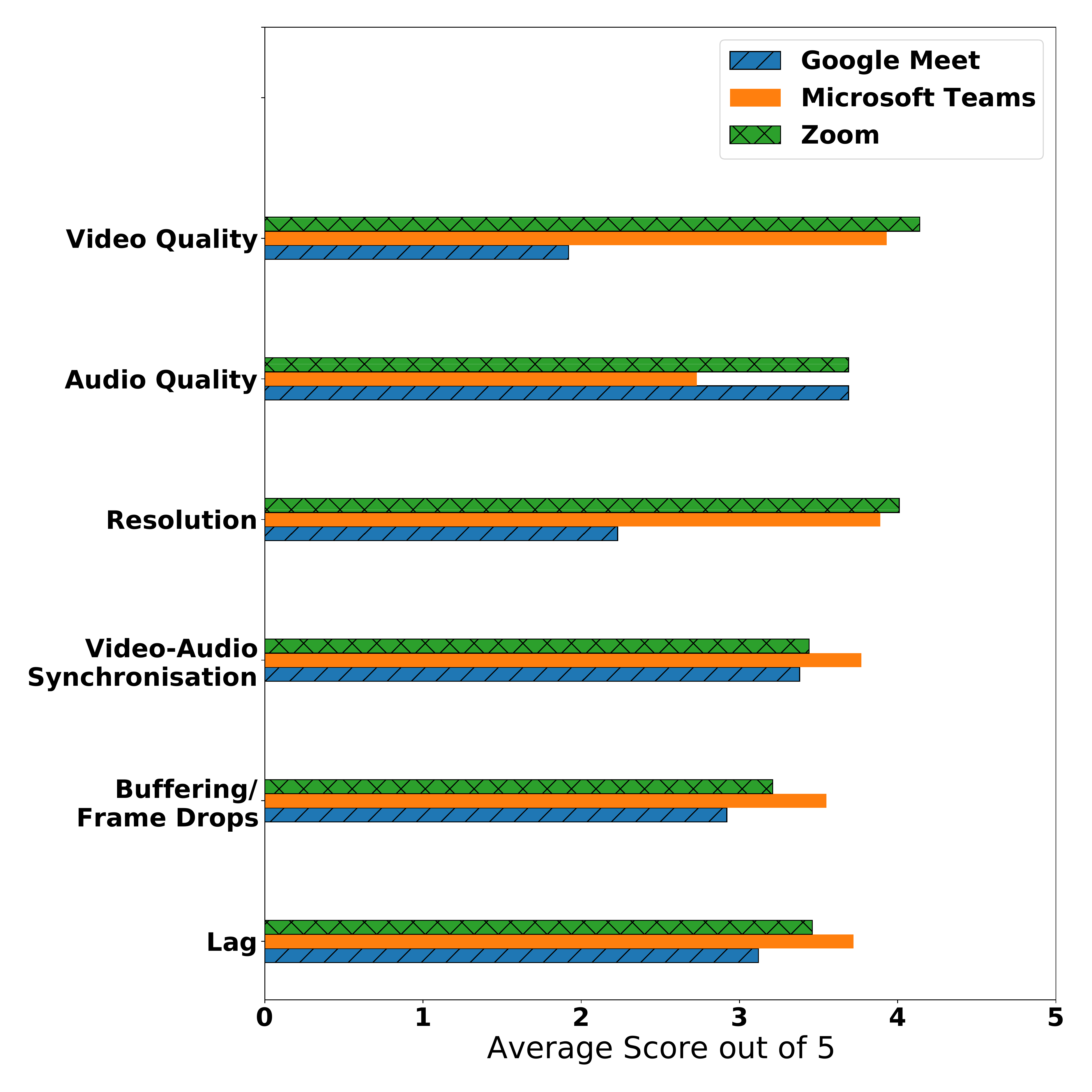}
    \caption{Qualitative Analysis Responses for 4G mobile Internet. The figure shows average score out of 5. Microsoft Teams and Zoom had a considerably better video quality and resolution than Google Meet. Google Meet and Zoom continues to provide a better Audio quality. However, Microsoft Teams provides better video-audio synchronization and lag free streaming.}\label{fig:subjective_4g}
\end{figure}

We show a qualitative comparison of the three apps in the qualitative analysis in Figures \ref{fig:subjective_bb} and \ref{fig:subjective_4g}.
For both wired broadband over WiFi and 4G mobile Internet connections, Zoom delivered the best video quality, while audio quality was comparable to Google Meet. Zoom provided the best overall performance for both the backhaul networks.

The Video Quality of Zoom is better for both wired broadband and 4G mobile Internet, followed by Microsoft Teams and Google Meet.
This corroborates with the higher SSIM and PSNR values.
Google Meet and Zoom have a better Video-Audio Synchronization as it has a lower standard deviation in the IPATs than that of Microsoft Teams in a wired broadband connection.
While for mobile 4G Internet, it is the reverse.
Microsoft Teams has a better Video-Audio synchronization in the 4G mobile Internet connection.
We see from the quantitative analysis that Google Meet reduces the Bandwidth by almost 40\% while on 4G mobile Internet. However, this affects the performance of Google Meet in the qualitative analysis for 4G mobile Internet. The subjects report a drop in Video Quality and Resolution for Google Meet while on 4G Internet.
Google Meet and Zoom have better Audio Quality than Microsoft Teams for both wired broadband and 4G mobile Internet. Users reported a consistent performance on Zoom for both WiFi and 4G. We see it in both the qualitative as well as quantitative analysis.

%

\section{Related Work}
\label{sec:related}
In this work, we conduct network and systems measurements using client-side network measurements and quantitative user studies for three popular video conferencing platforms.
In this section, we review some of the existing work and situate this paper in the context of the previous work.

\subsection{Networks and Systems Measurement Techniques}
Researchers have used client-side measurement techniques to understand closed systems' network structures and performance. Pevec \cite{youtube-measurements} measures YouTube traffic and performance based on the client-side measurements. Guha and Daswani \cite{skype-1} described a study to understand the P2P and supernode structures in the Skype network. Hoífeld and Binzenhöfer \cite{skype-2} conducted QoS and QoE evaluations for Skype over UMTS networks.
The QoS is analogous to our quantitative measurements, while QoE is analogous to our qualitative measurements.
Similar to our work, they use MOS (Mean Opinion Score) to estimate the QoE, throughput, IPAT, and packet loss to evaluate the QoS. In this work, we conduct these QoE and QoS measurements for two video conferencing apps over broadband and 4G mobile Internet. 
The authors in their paper \cite{10.1145/3487552.3487842} primarily measure how bandwidth varies depending on the number of users, while we are looking at how QoS and QoE vary for different backhaul networks.

Minimum download speed, average throughput for packet data, latency, drop rate, and successful data transmission percentage are optimized for 4G mobile Internet connections \cite{TRAI2018white}. Other researchers reported that network delay, jitter, and packet loss rate were also essential for a good QoE \cite{ning-2017}. These parameters served as a benchmark for acceptable quality over mobile Internet networks. Other researchers have focused on QoS and QoE measurements for multimedia applications over networks. Jiang and Saadawi \cite{rt-video-qos} described an approach for quality control on multimedia networks using QoE and QoS parameters. Similar to our approach, they used structural similarity to estimate QoE. Morshedi and Noll \cite{pqos-wifi} perform perceived QoS measurement in the WiFi access points in customers' premises. They used machine learning to develop models using which ISPs may estimate the QoS on their networks. On similar lines, Claypool and Tanner \cite{effects-of-jitter} measured the effects of jitter on the perceptual quality of the video. They found that jitter had the same impact on perceptual quality as packet losses. They discussed methods to quantify video quality objectively. 
Seshadrinathan and Bovik \cite{seshadrinathan-2010A} stated that any video quality analysis must be weighted based on Spatio-temporal distortion since video quality depends not only on resolution but also on the smoothness of playback. Researchers have also presented various metrics, and their efficacy differed with respect to the challenges faced in using them for judging video and network quality \cite{bouraqia-2020}. In this work, we use PSNR and SSIM to compare the received audio and video quality, similar to the work by Kotevski and Mitrevski \cite{kotevski-2010}.

\subsection{User Studies for User Experience}
Researchers have used user studies to develop algorithms to optimize user experience given a network scenario. Hossfeld et al. \cite{subjective-qoe} conducted user surveys to design an appropriate objective function for a Mixed Integer Linear Programming-based algorithm to benchmark the theoretical optimum for HAS (HTTP Adaptive Streaming) services. Qualitative user experience metrics were correlated with quantitative metrics to improve the overall quality of service by ISPs. The primary metrics for a smooth video conferencing experience were the stability of the connection and increased bandwidth. In this work, we use a 5-point Likert scale to conduct user studies to compare the two platforms we focus on in this study. 
While the paper \cite{10.1145/3487552.3487847} primarily studies lag experienced by different users in the USA and Europe over broadband, we also measure impact of using (a) 4G networks in India and (b) different combinations of using camera and mic on both, QoS and QoE.

Chen et al. \cite{user-satisfaction} conducted a user satisfaction study on online education platforms and found that the users' factors have no direct influence on user satisfaction. Instead, the availability of the platform plays an important role. Chen et al. \cite{covid-comments-analysis} analyzed comments on video conferencing platforms in China before and after the pandemic to understand the impact of the pandemic on the user experience with online education platforms. They found that the priorities and requirements of the users changed after the platforms started to be used for education during the pandemic. In our work, we augment a Networks and Systems analysis of three major video conferencing platforms with qualitative user studies to help determine the effect of the networks and systems characteristics on the user experience.

\subsection{Studies on Online Education over Video Conferencing Platforms}

Several researchers have documented the experience of online schooling and education during the COVID-19 pandemic. Our work in this domain \cite{cscw-blind} shows that students, families, and teachers faced several socio-cultural and technological barriers in access to online education. Chakraborty et al. \cite{covid-opinion} conducted a survey on online education during the pandemic with 358 respondents and found that the students felt physical classroom teaching works much better and that the students feel that the online classes are affecting their physical and social lives. In this work, we are evaluating only the technological aspects of the video conferencing platforms and not the socio-cultural aspects, such as familial support, mental and physical health, etc.

Our work is unique as it is the first one that evaluates the current video conferencing apps' quantitative and qualitative performance at the client-end on different networks with varied setups of camera and microphone.
We correlate the metrics so that we learn how to improve the apps in the future.


\section{Conclusion}
\label{sec:conclusion}

We find significant differences in how three video conferencing apps handled video and audio on two backhaul networks with different speeds.
Microsoft Teams uses approximately 10\% higher payloads than Google Meet and Zoom for wired broadband. This enables Microsoft Teams to provide video with higher resolution than Google Meet.
We see that Microsoft Teams has a marginally higher SSIM when the camera is switched OFF and has a considerably higher SSIM when the camera is switched ON as compared to Google Meet.
This results in consistently better video quality for Microsoft Teams than Google Meet.
Zoom is able to provide the same or even better video quality with significantly lesser payload.

A low SSIM value when the camera is switched ON suggests that Google Meet compresses the screen-sharing video to a more significant extent to compensate for the added payload when the camera is switched ON.
Google Meet achieves the compression by adjusting the luminescence of the video.
We observe that the ‘Y’ value for Google Meet is significantly lower than that of Microsoft Teams and Zoom when both the microphone and camera are switched ON, but the ‘U’ and ‘V’ values are comparable. The lower ‘Y’ values indicate that Google Meet compromises on the luminescence of the video to save bandwidth when the camera is switched ON. 

Microsoft Teams does not transmit packets at a steady rate, as seen from its higher standard deviation in IPAT, which is almost twice that of Google Meet and Zoom when the microphone and camera are switched OFF. This results in poor audio-video synchronization.
Although Microsoft Teams handles video better, it does not handle the audio in the same manner.
It compresses the audio to a greater extent by truncating the higher frequency notes. The bitrate of Microsoft Teams was substantially lower than that of Google Meet, and it also down-converted the audio from a stereo channel to a mono channel.
While Zoom gives a little higher bitrate than Google Meet, the latter provides more channels and therefore gives better audio quality.

Microsoft Teams and Zoom behave in the same manner on mobile 4G Internet as it does on wired broadband
, including their resource consumption.
However, Google Meet restricts its bandwidths while on 4G mobile Internet.
This further reduces its performance on video quality.
There is a considerable difference between Google Meet's video quality compared to the other two.
We observe that Google Meet's rate of packets is not as steady as it is for wired broadband over WiFi. 
We see higher buffering/frame drops for Google Meet now with higher memory consumption.
For audio, all three apps performed the same as wired broadband.
Therefore, Google Meet's audio continues to be better.

Our qualitative analysis confirms our quantitative analysis. Zoom and Microsoft Teams provide better video quality for both wired broadband over WiFi and 4G mobile Internet connections. On the audio front, Google Meet improves its quality as compared to the video front.
In a nutshell, if the audio is more important for the user, Google Meet and Zoom, in that order, are better options on wired broadband over WiFi.
Google Meet uses more memory, though.
While if the video is more important, Microsoft Teams and Zoom are better options.
Microsoft Teams needs more CPU for the task.
A user interested in saving the bandwidth needs to select Google Meet, especially when on 4G mobile Internet.
An ideal app would be one with Google Meet's audio and data payloads, Microsoft Team's IPAT, and Zoom's video quality.
The CPU utilization would be of Google Meet or Zoom, and memory consumption would be of Microsoft Teams.

\section{Future Work}
\label{sec:future}
We conducted our measurements on laptop-class machines. 
Many people use their smartphones, especially while they are moving.
The OS, these days, provides a setting to change its behavior in handling the camera. 
The video conferencing apps may also behave differently while on smartphones.
There is a need to execute similar experiments on smartphones to see how does app performs there.

While we considered three apps, there are other popular apps, e.g., Skype, Whatsapp, etc. 
Some of these apps do not restrict themselves to using the client-server paradigm.
They use the peer-to-peer paradigm or a combination of both paradigms.
They presumably have different behaviors when handling
video, audio, and working on different backhaul networks.
\bibliographystyle{ACM-Reference-Format}
\bibliography{refs}

\end{document}